 \documentclass[10pt,preprint]{aastex}

\usepackage{mathrsfs}

\newcommand{\LCDM}{$\Lambda$CDM }
\newcommand{\atl}{ATLAS$^{\rm 3D}$ }

\shorttitle{Modeling Nearly Spherical Pure-Bulge Galaxies with a $M_\star/L$ Gradient}
\shortauthors{Chae et al.}

\begin{document}


\title{Modeling Nearly Spherical Pure-Bulge Galaxies with a Stellar Mass-to-Light Ratio Gradient under the $\Lambda$CDM and MOND Paradigms: I. Methodology, Dynamical Stellar Mass, and Fundamental Mass Plane}


\author{Kyu-Hyun Chae}
\affil{Department of Physics and Astronomy, Sejong University, 
 209 Neungdong-ro Gwangjin-gu, Seoul 05006, Republic of Korea}
\email{chae@sejong.ac.kr}

\author{Mariangela Bernardi}
\affil{Department of Physics and Astronomy, University of Pennsylvania,
209 South 33rd Street, Philadelphia, PA 19104, USA}
\email{bernardm@sas.upenn.edu}

\and

\author{Ravi K. Sheth}
\affil{Department of Physics and Astronomy, University of Pennsylvania,
209 South 33rd Street, Philadelphia, PA 19104, USA}
\email{shethrk@physics.upenn.edu}




\begin{abstract}
We carry out spherical Jeans modeling of nearly round pure-bulge galaxies selected from the ATLAS$^{\rm 3D}$ sample. Our modeling allows for gradients in the stellar mass-to-light ratio ($M_\star/L$) through analytic prescriptions parameterized with a ``gradient strength'' $K$ introduced to accommodate any viable gradient. We use a generalized Osipkov-Merritt model for the velocity dispersion (VD) anisotropy. We produce Monte Carlo sets of models based on the stellar VD profiles under both the $\Lambda$CDM and MOND paradigms. Here, we describe the galaxy data, the empirical inputs, and the modeling procedures of obtaining the Monte Carlo sets. We then present the projected dynamical stellar mass, $M_{\rm \star e}$, within the effective radius $R_{\rm e}$, and the fundamental mass plane (FMP) as a function of $K$. We find the scaling of the $K$-dependent mass with respect to the ATLAS$^{\rm 3D}$ reported mass as: $\log_{10} \left[M_{\star{\rm e}}(K)/M_{\star{\rm e}}^{\rm A3D} \right] = a' + b' K$ with $a'=-0.019\pm 0.012$ and $b'=-0.18\pm 0.02$ ($\Lambda$CDM), or $a'=-0.023\pm 0.014$ and $b'=-0.23\pm 0.03$ (MOND), for $0\le K < 1.5$. The FMP has coefficients consistent with the virial expectation and only the zero-point scales with $K$. The median value of $K$ for the ATLAS$^{\rm 3D}$ galaxies is $\langle K\rangle =0.53^{+0.05}_{-0.04}$. We perform a similar analysis of the much larger SDSS DR7 spectroscopic sample.  In this case, only the VD within a single aperture is available, so we impose the additional requirement that the VD slope be similar to that in the ATLAS$^{\rm 3D}$ galaxies.  Our analysis of the SDSS galaxies suggests a positive correlation of $K$ with stellar mass.

\end{abstract}





\section{Introduction}

Galaxies are fascinating objects in their own right as basic building blocks of the universe, but they also provide crucial laboratories for cosmology and fundamental physics. A main goal of cosmology is to understand how galaxies form and evolve \citep{Mo10} and observed stellar kinematics in galaxies \citep{BT} provide evidence for unknown matter, referred to as dark matter (DM), or a new nature of gravity \citep{FM}.

Galaxies exhibit great varieties in morphological appearance, constituents, kinematics, and dynamics, etc. From the viewpoint of kinematics and dynamics, galaxies can be broadly divided into two classes: rotationally-supported versus pressure(dispersion)-supported. In reality, most galaxies have multiple components often containing both rotating and non-rotating components. For example, most spiral or disk galaxies contain non-rotating central bulges and most early-type (i.e.\ elliptical and lenticular) galaxies contain rotating disks \citep{Cap16}.  

In recent studies, it has been shown that rotating galaxies of various types satisfy Kepler-like kinematic laws such as the baryonic Tully-Fisher relation \citep{McG05} and the mass discrepancy-acceleration relation (or, radial acceleration relation) \citep{McG04,MLS,Lel}. Considering the potentially profound physical implications of the radial acceleration relation (see, e.g., Famaey, Khoury \& Penco 2018), it is interesting to explore whether there exists a universal radial acceleration relation for all galaxies, including non-rotating galaxies. In this respect, pure-bulge galaxies without rotating disks are of interest. These galaxies are also interesting for galactic astrophysics, e.g., as they may have been produced by the merging of galaxies, including rotating galaxies.

In this and companion/subsequent papers we analyze pure-bulge galaxies that also appear nearly round in projection. These galaxies form an extreme subset that is clearly distinct from other galaxies containing rotating components. One goal of studying these galaxies is to obtain an independent radial acceleration relation to compare with that of rotating galaxies and test theories of DM and gravity. The other goal is to model nearly spherical galaxies with a stellar mass-to-light ratio ($M_\star/L$) gradient included, as motivated from recent empirical findings \citep{MN,vD17,Son} and quantify the effects of a $M_\star/L$ gradient on dynamical estimates of stellar masses, VD anisotropy, and galactic structure.

We use 24 \atl galaxies with observed VD maps and 4201 SDSS galaxies with measured aperture VDs. We obtain Monte Carlo sets of models through the spherical Jeans equation based on the VD maps (or aperture VDs) and other empirical inputs under both the Lambda cold DM (\LCDM; see, e.g., Mo, van den Bosch \& White 2010) and modified Newtonian dynamics (MOND; Milgrom 1983) paradigms. Here we describe the data, the required empirical inputs and the modeling procedures. We then discuss dynamical estimate of stellar masses and the fundamental mass plane (FMP) in this paper. The radial acceleration relation based on the Monte Carlo sets is discussed in \cite{CBS17} while the effects of $M_\star/L$ gradient on VD anisotropy, stellar, DM and total mass profiles, DM mass fraction, and MOND interpolating function (IF) are discussed in subsequent publications. 

\section{Framework and Data}

We use strictly pure-bulge (i.e.\ spheroidal), mostly round (i.e.\ nearly spherical) galaxies selected from the ATLAS$^{\rm 3D}$ sample \citep{Cap11} and the SDSS DR7 spectroscopic sample \citep{DR7}. For our selected galaxies, the baryonic mass is primarily stellar -- gas and dust are negligible.  We carry out kinematic Jeans analyses based on three key pieces of data/information:
\begin{enumerate}
 \item The observed projected light distribution (i.e.\ surface brightness) $I(R)$, with its deprojected volume distribution $\rho_{\rm L}(r)$.
 \item The stellar mass-to-light ratio ($M_\star/L$), $\Upsilon_\star(R)=\Sigma_\star(R)/I(R)$, where $\Sigma_\star(R)$ denotes the projected stellar mass density.
 \item Line-of-sight (LOS) stellar velocity dispersions (VDs) at multiple radii, i.e.\ the radial profile $\sigma_{\rm los}(R)$ for \atl galaxies, or the mean value within an aperture radius $R_{\rm ap}$, $\sigma_{\rm ap}=\langle\sigma_{\rm los}\rangle(R=R_{\rm ap})$ (see below for details) for SDSS galaxies.
\end{enumerate}

\subsection{Stellar Mass-to-light Ratio Gradient}

We use an empirically motivated stellar mass-to-light ratio radial gradient given by
\begin{equation}
\frac{\Upsilon_\star (R/R_{\rm e})}{\Upsilon_{\star 0}} =\max\left\{1+ K\left[A - B (R/R_{\rm e})\right],1\right\},
 \label{eq:MLgrad}
\end{equation}
where $(A,~B)=(2.33,~6.00)$ are derived by \cite{Ber18} for the recently observed gradient \citep{vD17}. The parameter $K$ introduced here describes the gradient ``strength'':  $K=0$, $1$, and $0.555$ correspond, respectively, to no gradient, the strong gradient \citep{vD17}, and an intermediate gradient, i.e.\ the {\bf Salp$^{\rm In}$-Chab$^{\rm Out}$} model in which the stellar initial mass function (IMF) varies from the heavy \cite{Salp} IMF in the center to the light \cite{Chab} IMF in the outer region \citep{Ber18}. We consider the range $0\le K < 1.5$ to encompass all likely possibilities.

\subsection{Spherical Jeans Equation}

For a spherical galaxy with $\Sigma_\star(R)=\Upsilon_\star(R/R_{\rm e}) I(R)$ from which the baryonic volume density $\rho_{\rm B}(r)$ [$=\rho_\star(r)$ here] is obtained by deprojection, we solve the spherical Jeans equation \citep{BT}
\begin{equation}
\frac{d[\rho_{\rm B}(r) \sigma_{\rm r}^2(r)]}{dr} 
+ 2 \frac{\beta(r)}{r} [\rho_{\rm B}(r) \sigma_{\rm r}^2(r)]
= - \rho_{\rm B}(r) a(r),
 \label{eq:Jeans}
\end{equation}
for the radial stellar VD $\sigma_{\rm r}(r)$. Here, $a(r)$ is the gravitational acceleration (see \S 2.4). We assign a central black hole using a recent result \citep{Sag}. In Equation~(\ref{eq:Jeans}) the anisotropy parameter $\beta(r)=1 - \sigma_{\rm t}^2(r)/\sigma_{\rm r}^2(r)$ links the radial VD with the tangential VD $\sigma_{\rm t}(r)\equiv\sqrt{[\sigma_{\theta}^2(r)+\sigma_{\phi}^2(r)]/2}$. We consider radially constant anisotropies and also radially varying anisotropies given by
\begin{equation}
  \beta_{\rm gOM}(r)=\beta_0 + (\beta_\infty - \beta_0) \frac{(r/r_a)^2}{1+(r/r_a)^2},
 \label{eq:gOM}
\end{equation}
which varies smoothly from a central value $\beta_0$ to $\beta_\infty$ at infinity. We refer to this model as a generalized Osipkov-Merritt (gOM) model \citep{BT} since the combination of $\beta_0=0$ and $\beta_\infty=1$ corresponds to the Osipkov-Merritt model \citep{Osip,Merr}.

The observable LOSVD of stars at a projected radius $R$ on the sky, $\sigma_{\rm los}(R)$, is then given by \citep{BM82}
\begin{equation}
 I(R)\sigma_{\rm los}^2(R)=2 \int_{R}^{\infty}
 \rho_{\rm L}(r) \sigma_{\rm r}^2(r) \left[ 1 - \frac{R^2}{r^2} \beta(r) \right]
 \frac{rdr}{\sqrt{r^2-R^2}},
\label{eq:losvd}
\end{equation}
where we use light (rather than mass) densities $\rho_{\rm L}(r)$ and $I(R)$ appropriate for the case of radially varying $\Upsilon_\star(R)$ \citep{Ber18}. When aperture VDs are only available (but, radial profiles of LOSVDs are not), as is the case for SDSS galaxies, we will work with the light-weighted mean value of LOSVDs within a circular aperture:  
\begin{equation}
  \langle \sigma_{\rm los}\rangle (R) \equiv
  \frac{ \int_0^{R} I(R') \sigma_{\rm los}(R') R' dR'}{\int_0^{R} I(R')  R' dR'},
  \label{eq:VDap}
\end{equation}
where $\sigma_{\rm los}(R')$ is given by Equation~(\ref{eq:losvd}) and $I(R')$ is the surface brightness distribution.

\subsection{Integral Solution of the Spherical Jeans Equation}

For a given baryonic (stellar) mass density $\rho_{\rm B}(r)$ with a model for gravitational acceleration $a(r)$ under the \LCDM or the MOND paradigm, the solution of equation~(\ref{eq:Jeans}) for the radial stellar VD $\sigma_{\rm r}(r)$ can be
expressed, following appendix~B of \cite{Chae12}, as
\begin{equation}
\sigma^2_{\rm r}(r)=
   \int_r^\infty \frac{\omega(t)}{\omega(r)}  
   \frac{\rho_{\rm B}(t)}{\rho_{\rm B}(r)} a(t) dt,         
  \label{eq:sigint}
\end{equation}
where $\omega(r)=\exp\left[\int^r(2\beta(r')/r')dr'\right]$ for an anisotropy
$\beta(r')$ (Equation~\ref{eq:gOM}). This integral can be evaluated numerically. Typically it is sufficient to replace the infinite upper bound with $\sim 30$~$R_{\rm e}$. Examples of using this integral equation for the \LCDM and the MOND paradigm can be found in \cite{CBK} and \cite{CG}.

\subsection{Gravitational Acceleration}

For a galaxy with (observationally inferred) baryonic mass volume density $\rho_{\rm B}(r)$ the gravitational acceleration at $r$ depends on the assumed paradigm, i.e.\ \LCDM or MOND. Under the \LCDM paradigm we assume that the baryonic mass distribution is embedded in an extended DM mass distribution $\rho_{\rm DM}(r)$ (see \S 2.5). Then, the gravitational acceleration is given by
\begin{equation}
a(r) = G \frac{M_{\rm B}(r)+M_{\rm DM}(r)}{r^2},
 \label{eq:aDM}
\end{equation}
where $M_{\rm B}(r)$ and $M_{\rm DM}(r)$ are the integrated masses within $r$ for $\rho_{\rm B}(r)$ and $\rho_{\rm DM}(r)$, respectively.

Under the MOND paradigm the gravitational acceleration is given by
\begin{equation}
a(r) = f \left(\frac{a_{\rm B}(r)}{a_0}\right) a_{\rm B}(r)\hspace{1ex}{\rm with}\hspace{1ex} a_{\rm B}(r) = G \frac{M_{\rm B}(r)}{r^2},
 \label{eq:aMD}
\end{equation}
 where $a_{\rm B}(r)$ is the Newtonian acceleration predicted by the distribution of baryons. In Equation~(\ref{eq:aMD}) $f(x)$ is known as an IF (see \S 2.6) and $a_0$ is a free parameter known as the critical (or, characteristic) acceleration fitted to the data. The IF satisfies the limiting behaviors $f(x)\rightarrow 1$ for $x\equiv a_{\rm B}/a_0 \gg 1$ and $f(x) \rightarrow \sqrt(a_0/a_{\rm B})$ for $x \ll 1$, which means that only below $a_0$ does the empirical acceleration $a(r)$ deviate substantially from the Newtonian prediction $a_{\rm B}(r)$.

\subsection{Empirical Constraints on Dark Halos}

Our analysis under the \LCDM paradigm assumes that a spherical dark halo embeds each pure-bulge galaxy under consideration. To describe the unknown DM mass distribution we use the NFW model \citep{NFW} suggested from $N$-body simulations of CDM, but with its density slope parameter allowed to vary from the CDM-only prediction, so that possible effects of galaxy formation and evolution \citep{Mo10} can be mimicked. A generalized NFW (gNFW) model is described by
\begin{equation}
  \rho_{\rm gNFW}(r) \propto
  r^{-\alpha} \left[1+c_{200}\left(\frac{r}{r_{200}}\right)\right]^{-3+\alpha},
 \label{eq:gNFW} 
\end{equation}
where inner slope $\alpha$ and concentration $c_{200}$ are allowed to vary from the NFW values $\alpha_{\rm NFW}=1$ and $c_{\rm NFW}$. In Equation~(\ref{eq:gNFW}), $r_{200}$ denotes the radius of the sphere within which the DM density is 200 times the cosmic mean matter density. For a halo at redshift $z$, the radius $r_{200}$ is given by
\begin{equation}
 r_{200}= \frac{162.7}{1+z} \left(\frac{M_{200}}{10^{12}{\rm M}_\odot/h}\right)^{1/3} h^{-1} \Omega_{\rm m0}^{-1/3}\hspace{1ex} {\rm kpc}, 
 \label{eq:r200} 
\end{equation}
where $M_{200}$ is the integrated mass of the halo within $r_{200}$, $h=H_0/100$~km~s$^{-1}$~Mpc$^{-1}$ is the normalized Hubble constant, and $\Omega_{\rm m0}=\rho_{\rm m0}/\rho_{\rm crit 0}=\rho_{\rm m0}/[3H_0^2/(8\pi G)]$ is the cosmic mean matter density normalized by the critical density  $\rho_{\rm crit 0}=3H_0^2/(8\pi G)$ at $z=0$.
\begin{figure}[h] 
\begin{center}
\setlength{\unitlength}{1cm}
\begin{picture}(8,8)(0,0)
\put(-0.5,-1.){\includegraphics{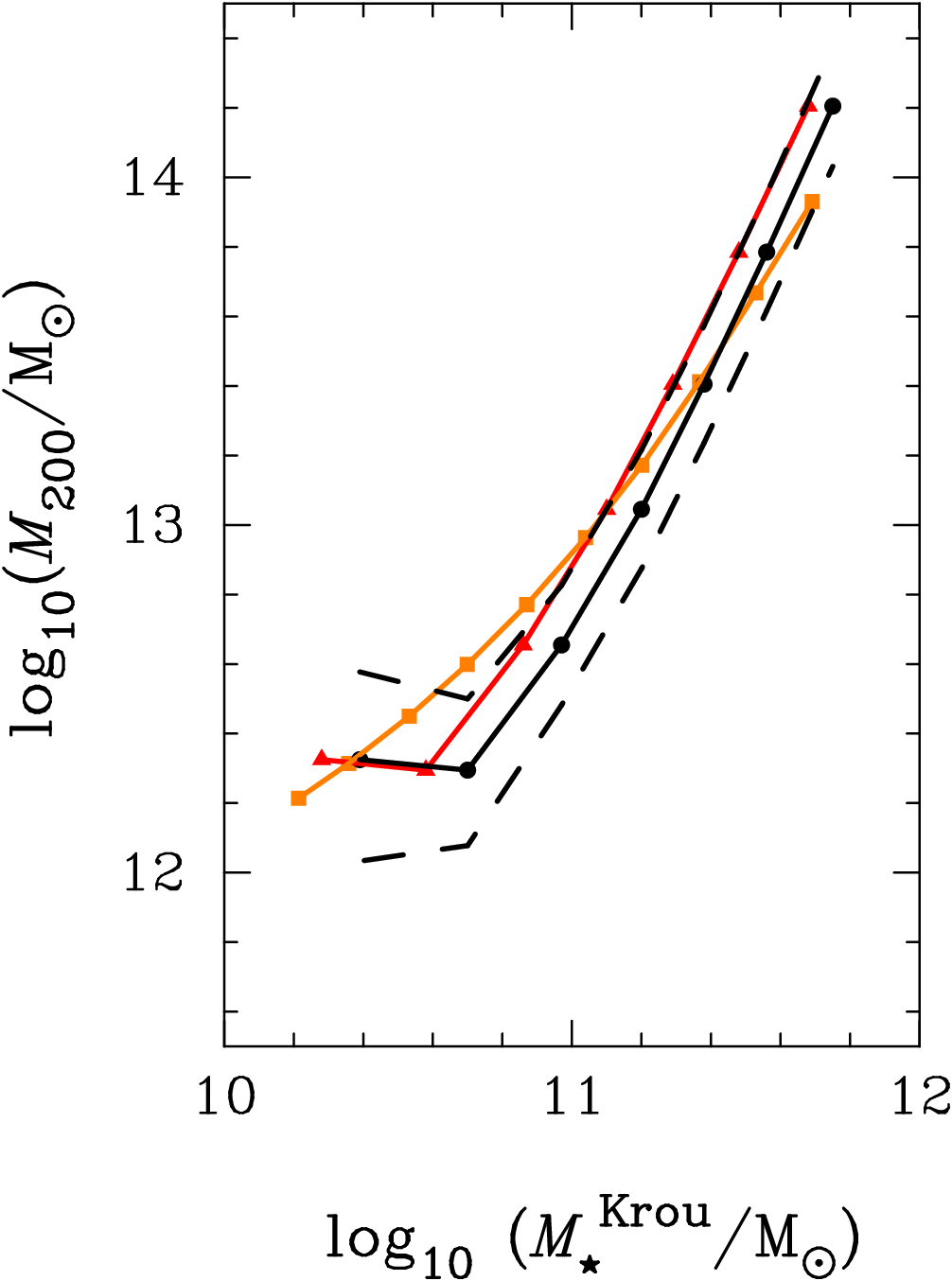}}
\end{picture}
\caption{Empirical stellar mass($M_\star^{\rm Krou}$)--halo mass($M_{200}$) relation used in this work under the DM paradigm. Here, $M_\star^{\rm Krou}$ denotes the stellar mass for the \cite{Krou} IMF, while $M_{200}$ denotes the mass within a spherical halo whose mean density is 200 times the cosmic mean matter density. The black solid and dashed curves represent, respectively, the weak-lensing relation by \cite{Man16} and our uncertainties (estimated including stellar mass uncertainties) based on their MPA-JHU stellar masses. The red triangles and curve are for their VAGC stellar masses. The orange squares and curve represent the relation from satellite kinematics by \cite{More}.}
\label{SMDM}
\end{center}
\end{figure}

The weak-lensing study by \cite{Man16} presents a numerical relation between $M_{200}$ and $M_\star^{\rm Krou}$ (their Table~B1) where $M_\star^{\rm Krou}$ is the total stellar mass of the galaxy estimated for the Kroupa \citep{Krou} IMF. Our fiducial choice of $M_\star^{\rm Krou}$ is their MPA-JHU value. The difference between their MPA-JHU and VAGC values is $\sim 0.09$~dex for $M_\star^{\rm Krou}\sim 10^{11}{\rm M}_\odot$. This uncertainty in $M_\star^{\rm Krou}$ and the measurement uncertainty of $\sim 0.05$~dex in $M_{200}$ give rise to a total uncertainty of $\sim 0.2$~dex in $M_{200}$ because $\log_{10}M_\star^{\rm Krou} \sim 2 \log_{10}M_{200}$ + const. Figure~\ref{SMDM} shows the relation with the estimated uncertainties. It also shows alternative relations, including that based on satellite kinematics \citep{More}, that are within the displayed uncertainties.

 We also use a recent $N$-body prediction of the $M_{200}$--$c_{\rm NFW}$ correlation \citep{DK}, which is consistent with weak-lensing constraint \citep{Man08}, and impose the constraint that the DM profile mimics the NFW profile for $r > 0.2 r_{200}$, consistent with weak-lensing empirical results \citep{Man08,Man16}. By doing so we are excluding unlikely wild combinations of $\alpha$ and $c_{200}$. For this we follow the simple procedure described in Section~3.3.1 of \cite{CBK}.

\subsection{MOND IF}

 For the MOND IF defined in Equation~(\ref{eq:aMD}) we use the family of models indexed by $\nu$, 
\begin{equation}
  f_\nu(x) = \left(\frac{1}{2}+\sqrt{\frac{1}{4}+\frac{1}{x^{\nu}}}\right)^{1/\nu},
 \label{eq:IFnu}
\end{equation}
where $\nu=1$ is the `simple' \citep{FB} and $\nu=2$ is the `standard' case \citep{Ken}. We consider $0< \nu \le 2$, which can encompass various possibilities suggested in the literature. See \cite{CBS17} for other functional forms.

\subsection{ATLAS$^{\rm 3D}$ Galaxies}

\subsubsection{Light Distribution, Mass-to-light Ratio Gradient, and Stellar Mass Distribution}

The \atl team \citep{Scot} describes the 2-dimensional distribution of light (i.e.\ surface brightness) using the so-called multi-Gaussian expansion (MGE) model \citep{EMB}:
\begin{equation}
  I(X,Y)=\sum_{j=1}^{N} I'_j \exp\left[-\frac{1}{2\sigma_j^2}\left(X^2 + \frac{Y^2}{{q'}_j^2}\right)\right],
 \label{eq:IMGE}  
\end{equation}
where parameters $I'_j$, $\sigma_j$, and $q'_j$ can be found from the \atl project website (http://www-astro.physics.ox.ac.uk/atlas3d/). The luminosity based on the MGE model is given by
\begin{equation}
L_{\rm MGE}=\sum_{j=1}^{N}2\pi I'_j \sigma_j^2 q'_j.  
 \label{eq:LMGE}  
\end{equation}
This is likely to be an underestimate of the true total luminosity because the expansion terms typically do not include light at large radii ($\gtrsim$ 2 -- 3 $R_{\rm e}$), and there is no extrapolation term to estimate light at large radii. For the same reason the half-light radius ($R_{\rm e,MGE}$) based on Equations~(\ref{eq:IMGE}) and (\ref{eq:LMGE}) is smaller than the true value $R_{\rm e}$. In fact, the ATLAS$^{\rm 3D}$ team reports $R_{\rm e}=1.35R_{\rm e,MGE}$ \citep{Cap13a}. We estimate the projected light within $R_{\rm e}$ using an MGE light distribution of $\approx 0.594 L_{\rm MGE}$, from which we estimate the total luminosity as $L_{\rm tot}\approx 1.188 L_{\rm MGE}$.

In this work we only consider nearly round pure-bulge galaxies under the spherical symmetry assumption. We thus circularize the observed light distribution with the substitution ($\sigma_j \rightarrow \sigma_j \sqrt{q'_j}$, $q'_j \rightarrow 1$), following, e.g.\ \cite{Cap13a}, as
\begin{equation}
I(R)=\sum_{j=1}^{N}I'_j\exp\left(-\frac{1}{2\sigma_j^2 q'_j} R^2\right), 
 \label{eq:IRMGE}  
\end{equation}
where $R=\sqrt{X^2+Y^2}$ is the radius on the sky. The integrated light within $R$ on the 2-dimensional plane is then given by
\begin{equation}
L^{\rm 2D}(R)= 2 \pi \sum_{j=1}^{N} I'_j \sigma_j^2 q'_j
\left[ 1 - \exp\left(-\frac{1}{2\sigma_j^2 q'_j}R^2 \right) \right]. 
 \label{eq:L2RMGE}  
\end{equation}

Deprojecting the 2-dimensional circular light distribution (Equation~\ref{eq:IRMGE}) yields 
\begin{equation}
\rho_{\rm L}(r)=\sum_{j=1}^{N}I'_j\sqrt{\frac{1}{2\pi \sigma_j^2q'_j}}
    \exp \left( -\frac{1}{2\sigma_j^2 q'_j } r^2 \right), 
 \label{eq:rhoLMGE}  
\end{equation}
where $r$ is the three-dimensional (rather than projected) radius. The integrated light within $r$ in the 3-dimensional space is then given by
\begin{equation}
L^{\rm 3D}(r)= 2 \pi \sum_{j=1}^{N} I'_j \sigma_j^2 q'_j
\left[ {\rm erf}\left(\sqrt{k_j}r\right) -2 \sqrt{\frac{k_j}{\pi}}r
  \exp\left(-k_j r^2 \right) \right], 
 \label{eq:L3MGE}  
\end{equation}
where $k_j \equiv 1/(2\sigma_j^2q'_j)$ and ${\rm erf}(x)$ is the error function.

\begin{figure}[h] 
\begin{center}
\setlength{\unitlength}{1cm}
\begin{picture}(7,7)(0,0)
\put(-2.5,7.5){\includegraphics{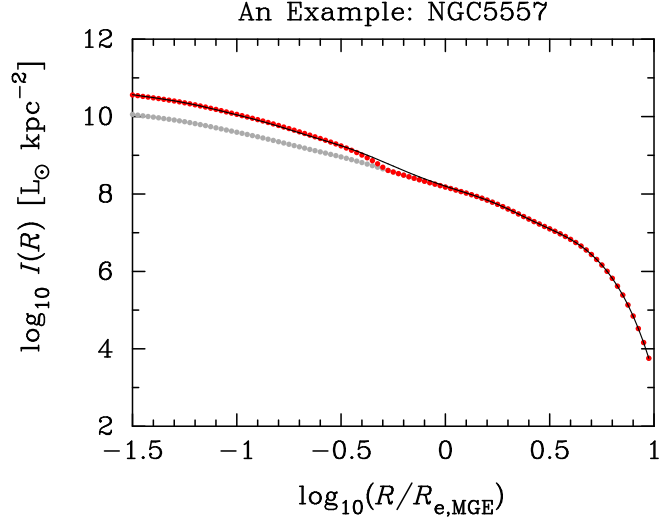}}
\end{picture}
\caption{Observed surface brightness distribution described by the MGE model (gray) and the effective distribution (red) defined by multiplying $\Upsilon_\star(R)/\Upsilon_{\star 0}$ (Equation~\ref{eq:MLgrad} with $K=1$) by the observed distribution. The black curve fitted to red dots is the modified MGE model (Equation~\ref{eq:ItilR}) with the modified coefficients defined by Equation~(\ref{eq:Itil}). }
\label{SBRMGE}
\end{center}
\end{figure}

If the stellar mass-to-light ratio were constant (denoted by $\Upsilon_{\star 0}$) throughout a galaxy, then the stellar mass density $\rho_\star(r)$ would simply be equal to Equation~(\ref{eq:L3MGE}) multiplied by $\Upsilon_{\star 0}$. For radially varying $\Upsilon_\star(R)$ the projected stellar mass density can be written by
\begin{equation}
\Sigma_\star(R) = \Upsilon_\star(R) I(R)
 =\Upsilon_{\star 0} \frac{\Upsilon_\star(R)}{\Upsilon_{\star 0}} I(R)
 = \Upsilon_{\star 0} \tilde{I}(R), 
 \label{eq:S2R}  
\end{equation}
where $\Upsilon_{\star 0}$ is a constant representing $M_\star/L$ for $R>0.4R_{\rm e}$ (cf.\ Equation~\ref{eq:MLgrad}) and we have defined an effective light distribution $\tilde{I}(R)\equiv\left[\Upsilon_\star(R)/\Upsilon_{\star 0}\right] I(R)$. For $\Upsilon_\star(R)/\Upsilon_{\star 0}$ given by Equation~(\ref{eq:MLgrad}), deprojection of $\Sigma_\star(R)$ [or $\tilde{I}(R)$] may be cumbersome. Instead of considering the direct deprojection, we rescale the coefficients $I'_j \rightarrow \tilde{I}'_j$ using a prescription given by
\begin{equation}
\tilde{I}'_j=I'_j\times {\rm max}\left\{1+K\left[-1+f_0+f_1\left(\sigma_j/R_{\rm e} \right)\right],1\right\}, 
 \label{eq:Itil}  
\end{equation}
where $K$ is the gradient strength defined in Equation~(\ref{eq:MLgrad}) and parameters $f_0$ and $f_1$ are fitted for each galaxy. For example, NGC~5557 has $f_0=3.745$ and $f_1=-6.594$ determined for $K = 1$. Then, the effective distribution $\tilde{I}(R)$ in Equation~(\ref{eq:S2R}) can be approximated by
\begin{equation}
\tilde{I}(R)=\sum_{j=1}^{N}\tilde{I}'_j\exp\left(-\frac{1}{2\sigma_j^2 q'_j} R^2\right). 
 \label{eq:ItilR}  
\end{equation}
Figure~\ref{SBRMGE} shows the observed light distribution $I(R)$, the effective distribution given by Equation~(\ref{eq:S2R}), and the distribution given by Equation~(\ref{eq:ItilR}) fitted with Equation~(\ref{eq:Itil}) for NGC~5557. One can see that the fitted distribution matches well the effective distribution for most radial range, but not near the transition radius $0.4 R_{\rm e}$. Actually, the fitted distribution even appears to be preferable, as it varies more smoothly in the transition region.

Based on the above, the projected stellar mass density can be approximated by
\begin{equation}
 \Sigma_\star(R) = \Upsilon_{\star 0} \tilde{I}(R)
  = \Upsilon_{\star 0} \sum_{j=1}^{N}\tilde{I}'_j\exp\left(-\frac{1}{2\sigma_j^2 q'_j} R^2\right), 
 \label{eq:S2Rtil}  
\end{equation}
where the $\tilde{I}'_j$ are the effective coefficients given by Equation~(\ref{eq:Itil}). The volume stellar mass density needed for the Jeans equation (Equation~\ref{eq:Jeans}) is then 
\begin{equation}
\rho_{\rm B}(r)=\Upsilon_{\star 0} \sum_{j=1}^{N}\tilde{I}'_j\sqrt{\frac{1}{2\pi \sigma_j^2q'_j}} \exp \left( -\frac{1}{2\sigma_j^2 q'_j } r^2 \right),  
 \label{eq:rhoBr}  
\end{equation}
and the baryonic (stellar) integrated mass within $r$ is given by
\begin{equation}
M_{\rm B}(r)= 2 \pi \Upsilon_{\star 0} \sum_{j=1}^{N} \tilde{I}'_j \sigma_j^2 q'_j
\left[ {\rm erf}\left(\sqrt{k_j}r\right) -2 \sqrt{\frac{k_j}{\pi}}r
  \exp\left(-k_j r^2 \right) \right]. 
 \label{eq:MBr}  
\end{equation}
Note that Equations (\ref{eq:rhoBr}) and (\ref{eq:MBr}) can be, respectively, obtained by multiplying Equations (\ref{eq:rhoLMGE}) and (\ref{eq:L3MGE}) by $\Upsilon_{\star 0}$ with the replacement of $I_j \rightarrow I_j'$ (Equation~\ref{eq:Itil}).

\subsubsection{Selection of Pure-bulge Galaxies and Their VD Profiles}

\begin{figure}[h] 
\begin{center}
\setlength{\unitlength}{1cm}
\begin{picture}(7,7)(0,0)
\put(-2.5,7.5){\includegraphics{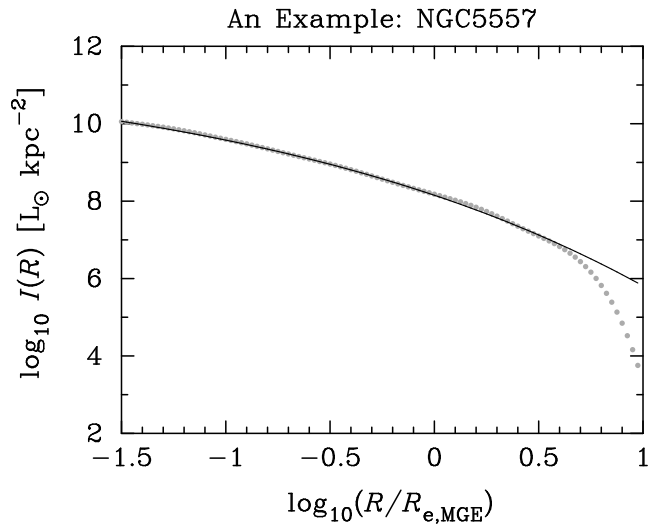}}
\end{picture}
\caption{Example of fitting the observed surface brightness distribution described by the MGE model (gray dots) with the S\'{e}rsec model (black curve) with the constraint $R_{\rm e}=1.35 R_{\rm e,MGE}$. Note that the \atl provided MGE distribution declines rapidly well outside $R_{\rm e}$ due to the absence of higher-order terms.}
\label{SBRMGEser}
\end{center}
\end{figure}

The vast majority of the 260 \atl galaxies possess some rotation \citep{Ems}. This means that only a small fraction of the \atl galaxies can be included in our selection of (little-rotating) pure-bulge galaxies, for which the spherical Jeans equation applies. Our selection of pure-bulge galaxies is based on a photometric analysis by \cite{Kra} who decomposed the observed light distribution into two components, a disk described by an exponential profile and a bulge described by a S\'{e}rsic profile for galaxies that do not possess bars. They report the relative amounts of light contained in the two components. We only take galaxies for which a disk is not detected at all or the measured light in the disk is less than 5\% of the total light, and the reported S\'{e}rsic index $n > 3$ (to make sure that the bulge is clearly different from the disk profile). This resulted in the following 27 galaxies: NGC~0661 ($n=5.4$), NGC~1289 (4.3), NGC~2695 (5.2), NGC~3182 (3.5), NGC~3193 (5.0), NGC~3607 (4.7), NGC~4261 (4.4), NGC~4365 (4.3), NGC~4374 (5.0), NGC~4406 (3.6), NGC~4459 (4.3), NGC~4472 (4.7), NGC~4486 (4.2), NGC~4636 (3.8), NGC~4753 (2.9), NGC~5322 (4.6), NGC~5481 (4.2), NGC~5485 (3.7), NGC~5557 (4.6), NGC~5631 (4.9), NGC~5831 (3.9), NGC~5846 (3.5), NGC~5869 (5.2), NGC~6703 (4.7) and NGC~0680 (7.6), NGC~4552 (7.3), NGC~5576 (8.3). The values provided in parentheses are our estimates of the S\'{e}rsic index by fitting the observed light distributions with fixed $R_{\rm e}=1.35 R_{\rm e,MGE}$. An example of such fits can be found in Figure~\ref{SBRMGEser}.  We also fitted these galaxies with a constraint $n < n_{\rm max}$ with $R_{\rm e}$ allowed to be free. In this case we find that the fitted $R_{\rm e}$ matches $1.35 R_{\rm e,MGE}$ only for $n_{\rm max}\le 5.5$. As $n$ gets larger than $5.5$, the fitted $R_{\rm e}$ becomes increasingly larger than $1.35 R_{\rm e,MGE}$. This raises potential problems in using $R_{\rm e}=1.35 R_{\rm e,MGE}$ for galaxies with $n > 5.5$ as we use $R_{\rm e}=1.35 R_{\rm e,MGE}$ throughout, particularly in deriving the FMP for pure-bulge galaxies. To avoid any biases we thus exclude the last three galaxies in the list above -- NGC0~680, NGC~4552, and NGC~5576 -- from our analysis. The final 24 pure-bulge galaxies have accurate light distributions up to $\sim 2$ -- $3 R_{\rm e}$ described by the MGE model and the well-understood $R_{\rm e}$ using the extrapolated S\'{e}rsic light profiles.

 Our photometrically defined pure-bulge \atl galaxies are nearly round, with a mean minor-to-major axis ratio of $\langle b/a \rangle=0.82$. Nearly round galaxies with a low disk-to-total light ratio (D/T) and large S\'{e}rsic index $n>3$ are expected to be little-rotating \citep{Kra}. Indeed, the selected 24 pure-bulges are overall consistent with kinematic identifications. Based on the angular momentum parameter $\lambda_{R_{\rm e}}$ within $R_{\rm e}$  (Table B1, \cite{Ems}), 16 (67\%) out of the 24 pure-bulges are classified as slow rotators. Our Jeans analysis is little affected whether we exclude the 8 kinematically fast rotators or not. We will use all 24 (photometrically identified) pure-bulges for our analyses. 

\begin{figure}[h] 
\begin{center}
\setlength{\unitlength}{1cm}
\begin{picture}(7,7)(0,0)
\put(-1.,-2.5){\includegraphics{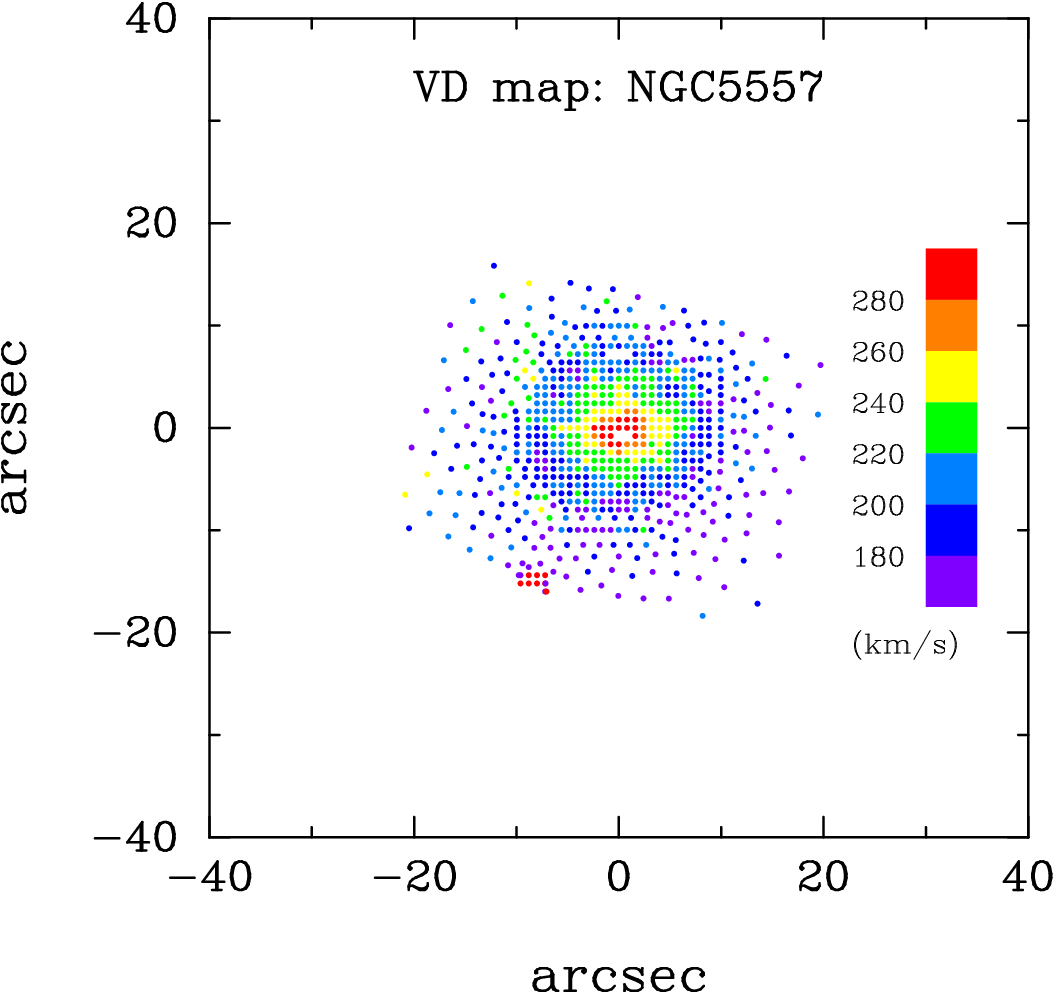}}
\end{picture}
\caption{Example of the two-dimensional map of observed LOSVDs that has an approximate circular symmetry.}
\label{VDmap}
\end{center}
\end{figure}

\begin{figure}[h] 
\begin{center}
\setlength{\unitlength}{1cm}
\begin{picture}(9,9)(0,0)
\put(-0.3,-0.5){\includegraphics{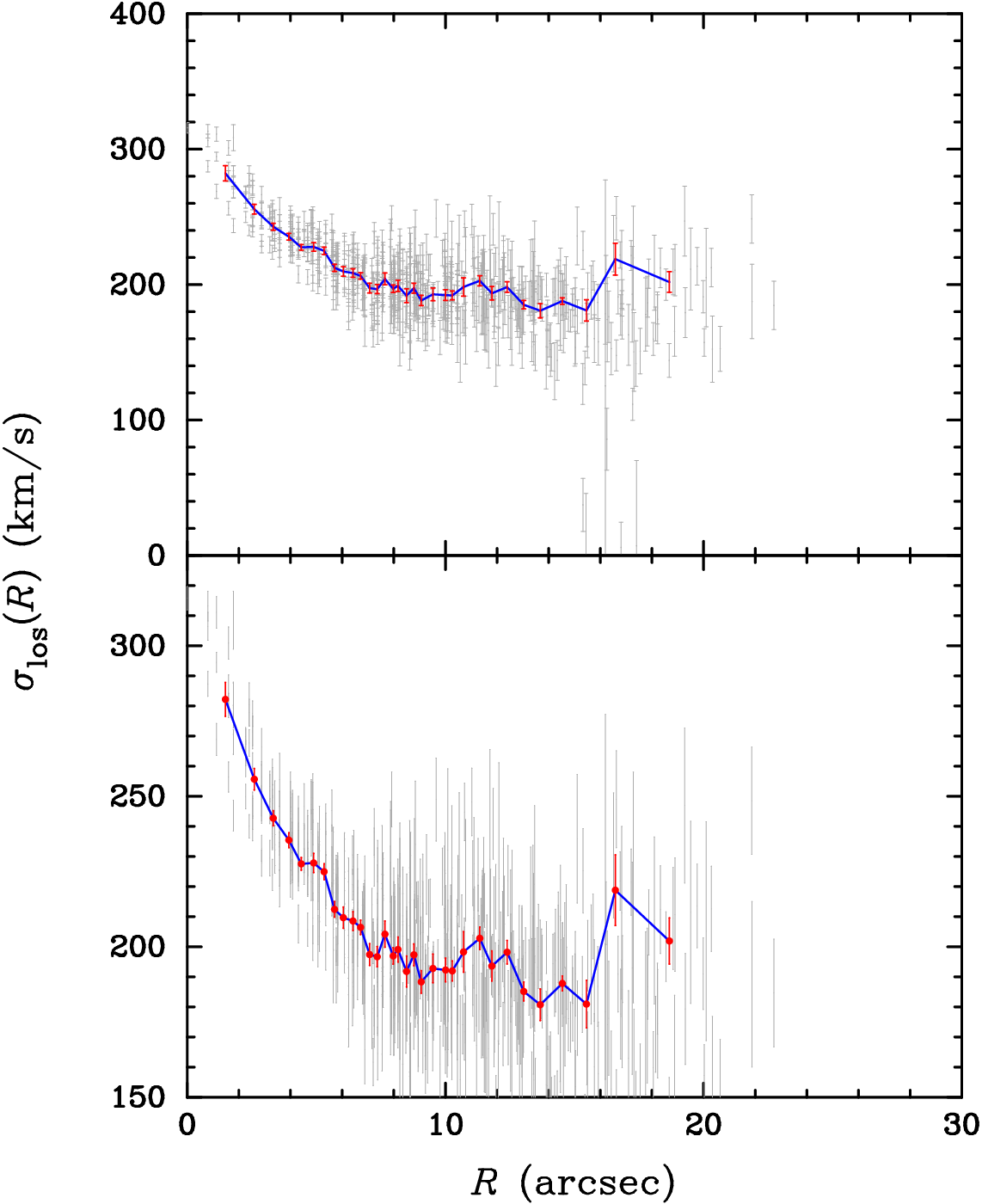}}
\end{picture}
\caption{Radial profile of LOSVDs constructed using the VD map of Figure~\ref{VDmap}. Radial bins are formed by concentric rings in the VD map so that each ring except for the last ring contains 21 VDs. The gray points and bars represent the measured VDs and uncertainties. The red points with error bars are the weighted means, with their uncertainties estimated by the 68\% scatters divided by $\sqrt(N-1)$. The bottom panel is a zoomed-in view magnified vertically. }
\label{VDprof}
\end{center}
\end{figure}

\begin{figure} 
\begin{center}
\setlength{\unitlength}{1cm}
\begin{picture}(10,18)(0,0)
\put(-2.8,-1.1){\includegraphics{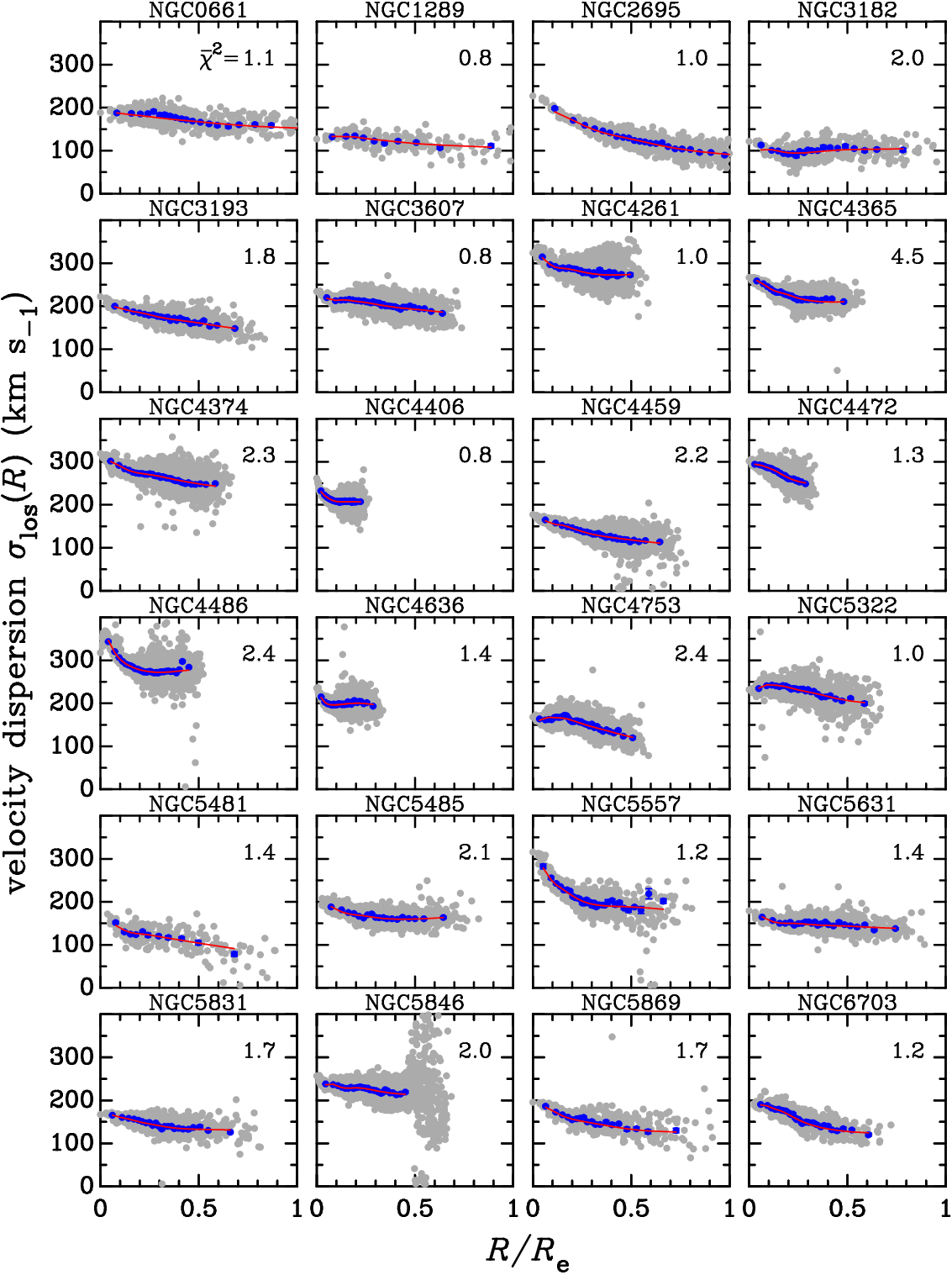}}
\end{picture}
\caption{Line-of-sight velocity dispersion profiles of 24 pure-bulge \atl galaxies. The gray points are the measured values on the sky distributed with radius. The blue points represent the weighted means in radial bins, each bin containing at least 21 measurements. Each red curve represents the prediction by the best-fit model for the galaxy with the gNFW DM halo (Equation~\ref{eq:gNFW}) and the gOM anisotropy (Equation~\ref{eq:gOM}). The best-fit reduced chi-squared $\bar{\chi}^2\approx \chi^2/N_{\rm bins}$ value is given in each panel. Modeling under the MOND paradigm gives similar fit results.}
\label{VDP_A3D}
\end{center}
\end{figure}

The \atl galaxies have detailed maps of LOSVDs up to $\sim 1R_{\rm e}$ which are exhibited in \cite{Cap13a}. An example is reproduced in Figure~\ref{VDmap} using the public data for NGC~5557 obtained from the project page. This figure contains 692 data points. We construct a radial VD profile using the data points. Figure~\ref{VDprof} exhibits the distribution of the measured VDs with radius for NGC~5557. To construct a VD profile the observed radial range is divided into radial bins so that each bin contains at least 21 VDs. Using the VDs in each bin we calculate the statistically weighted mean (or the median) and the 68\% scatter from which we estimate the uncertainty of the mean assuming the normal distribution in the bin. Figure~\ref{VDprof} shows the constructed VDP of NGC5557 for which each bin contains 21 data points except for the last bin. Figure~\ref{VDP_A3D} exhibits the VDPs for all 24 pure-bulge galaxies.

\subsection{SDSS Galaxies}

\subsubsection{SDSS Pure-bulge Galaxies: Sample Selection}

The final data release of SDSS I (DR7) provides light distributions and spectra for nearly one million galaxies \citep{DR7}. The enormous sample size is useful for addressing many science issues, including the radial acceleration relation and galactic structure. However, SDSS galaxy spectra were obtained using a fixed aperture radius of $\theta_{\rm ap} = 1.5$ arcsec. This means that the SDSS aperture VD $\sigma_{\rm ap}$ equals $\langle \sigma_{\rm los}\rangle (R_{\rm ap})$ of Equation~(\ref{eq:VDap}), where $R_{\rm ap}=\theta_{\rm ap}D$ and $D$ is the distance to each galaxy.  We convert from the measured redshift $z$ to the distance $D$ using a flat \LCDM cosmological model with the present matter density of $\Omega_{\rm m0}=0.3$ and the Hubble constant of $H_0 =70$~km~s$^{-1}$~Mpc$^{-1}$.

With $\sigma_{\rm ap}$ measured at a single scale only, rather than a VD profile $\sigma_{\rm los}(R)$ over a range of scales, $M_\star/L$ and DM distribution (or MOND IF) cannot be well constrained by the SDSS data alone. Therefore, we use an FMP derived from \atl pure-bulge galaxies (see \S 3.2) to provide an additional empirical constraint.  

We draw pure-bulge, round galaxies from the UPENN database of $\approx 0.7$ million galaxies \newline ({http://www.physics.upenn.edu/$\sim$ameert/SDSS$\underbar{\hspace{1ex}}$PhotDec/}) for which light distributions have been thoroughly analyzed \citep{Mee}. For most of these galaxies the aperture VDs have been reliably derived by two groups (MPA-JHU, Portsmouth) (see {http://www.sdss.org/dr12/spectro/galaxy/} and also \cite{ThoD} for the Portsmouth VDs). We select pure-bulge galaxies using flag bit 2 in the UPENN catalog and further require that galaxies have probabilities of being elliptical of greater than 70\% using an automated morphological classification \citep{HC} to exclude lenticular galaxies seen face-on by chance. To select round galaxies we require a minor-to-major axis ratio $ > 0.85$.  These are conservative cuts that are satisfied by only $\approx$11,000 galaxies.  Of these, we only retain those galaxies whose photometric measurements ($R_{\rm e}$ and radial light profile) and subsequent FMP-based estimate of stellar mass density are robust. For this we require SDSS $r$-band absolute magnitude $M_r > -23$ (to exclude too bright and large galaxies), stellar mass $M_{\star} > 10^{10.4}~{\rm M}_\odot$ (here $M_{\star}$ refers to the FMP-based value), and S\'{e}rsic index $3 < n < 5.5$, where the lower cut is to ensure light profiles well distinct from disk galaxies and the upper cut is to ensure that the S\'{e}rsic-based $R_{\rm e}$ matches the \atl reported $R_{\rm e}$. This upper cut matters because we are using the FMP relation that is derived using the \atl reported $R_{\rm e}$ values. Lastly, we use only those galaxies whose measured VDs have uncertainties smaller than $1/3$ of the values (but, typical uncertainties are small, just $0.02$--$0.04$~dex). Our final sample contains 4201 (4229) galaxies for MPA-JHU (Portsmouth) VDs. The typical difference between the two VDs is 0.02~dex (or 5\%) which shows a good agreement between the two. We will use the MPA-JHU VDs throughout, but with their formal errors multiplied by $\sqrt{2}$, considering the small difference with the Portsmouth VDs.
 
\subsubsection{SDSS Pure-bulge Galaxies: Statistical Properties and Implicit Stacking}

\begin{figure}[h] 
\begin{center}
\setlength{\unitlength}{1cm}
\begin{picture}(7,7)(0,0)
\put(-3.,7.5){\includegraphics{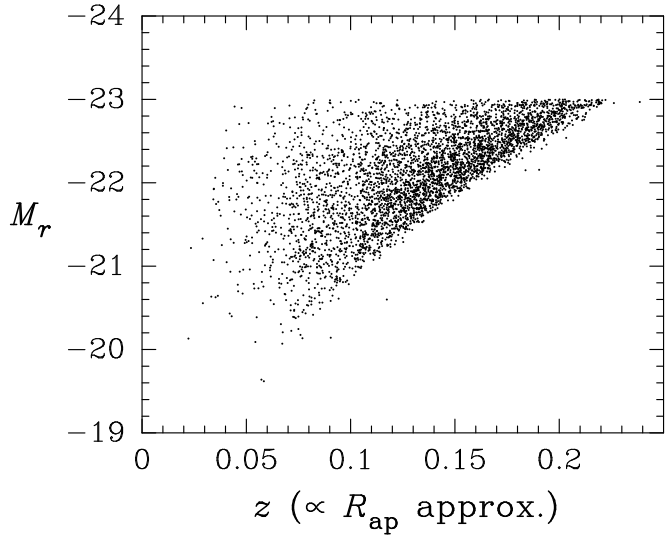}}
\end{picture}
\caption{Distribution of 4201 nearly round pure-bulge SDSS galaxies in the redshift($z$)-absolute magnitude($M_r$) plane, where $M_r$ is the absolute magnitude in the SDSS $r$-band. Because SDSS used a fixed spectroscopic aperture radius of $1.5$~arcsec, at fixed $M_r$ galaxies of different redshifts have aperture VDs within different projected aperture radii, $R_{\rm ap}$ that is approximately scaling linearly with $z$ for the shown redshift range.}
\label{zMr}
\end{center}
\end{figure}

Figure~\ref{zMr} shows the distribution of galaxies in the redshift($z$)-absolute magnitude($M_r$) plane. At low redshift ($z \lesssim 0.2$) the angular-diameter distance to a galaxy increases approximately linearly with $z$ so that the physical aperture radius $R_{\rm ap}$ for a fixed angular radius $\theta_{\rm ap}=1.5$~arcsec is approximately proportional to $z$. Thus, many galaxies of the same luminosity have different values of $R_{\rm ap}$ providing a stacked profile of  $\sigma_{\rm ap}(R_{\rm ap})$ at fixed $M_r$, something similar to the observed LOSVD profile $\sigma_{\rm los}(R)$ in \atl galaxies. We can construct a better stacked profile of $\sigma_{\rm ap}(R_{\rm ap})$ using additional galaxy parameters. For example, we can construct  $\sigma_{\rm ap}(R_{\rm ap})$ at fixed $M_r$, $R_{\rm e}$ and $n$. This stacking is implicit in our analysis based on SDSS galaxies. In other words, although each galaxy uses just one value of the aperture VD at $R_{\rm ap}$, those values can be attributed to certain stacked profiles.

\subsubsection{Double-Gaussian Model to Represent the Stellar Mass-to-light Ratio Gradient for the S\'{e}rsic Light Profile}

Unlike the MGE model in which its own parameters can be adjusted to accommodate the effect of a $M_\star/L$ gradient, the S\'{e}rsic light profile does not allow such a flexibility because it has only two parameters to describe the shape, i.e., $n$ and $R_{\rm e}$. It is convenient to represent the effect of a $M_\star/L$ gradient in the central region using an analytically tractable double-Gaussian model written as
\begin{equation}
I_{\rm cen}(R)=I_{\rm cen,1} \exp\left(-\frac{R^2}{2\sigma_{\rm cen,1}^2}\right)+I_{\rm cen,2} \exp\left(-\frac{R^2}{2\sigma_{\rm cen,2}^2}\right),   
 \label{eq:IcenR}  
\end{equation}
whose total luminosity is $L_{\rm cen}=2\pi (I_{\rm cen,1}\sigma_{\rm cen,1}^2 + I_{\rm cen,2}\sigma_{\rm cen,2}^2) $. The deprojected volume density is given by
\begin{equation}
\rho_{\rm L,cen}(r)=\frac{I_{\rm cen,1}}{\sqrt{2\pi \sigma_{\rm cen,1}^2}} \exp\left(-\frac{r^2}{2\sigma_{\rm cen,1}^2}\right)+\frac{I_{\rm cen,2}}{\sqrt{2\pi \sigma_{\rm cen,2}^2}} \exp\left(-\frac{r^2}{2\sigma_{\rm cen,2}^2}\right),   
 \label{eq:rhocenr}  
\end{equation}
and the integrated luminosity within $r$ is given by
\begin{equation}
L_{\rm cen}(r)= 2 \pi \sum_{j=1}^{2} I_{\rm cen,j} \sigma_{\rm cen,j}^2 
\left[ {\rm erf}\left(\frac{r}{\sqrt{2}\sigma_{\rm cen,j}}\right) - \sqrt{\frac{2}{\pi \sigma_{\rm cen,j}^2}}r  \exp\left(- \frac{r^2}{2\sigma_{\rm cen,j}^2} \right) \right]. 
 \label{eq:Lcenr}  
\end{equation}

For the light distribution described by a S\'{e}rsic light profile $I_{\rm Ser}(R)$ and the radially varying stellar mass-to-light ratio $\Upsilon_\star(R)$ (Equation~\ref{eq:MLgrad}), the projected stellar mass density (see Equation~\ref{eq:S2R}) can be approximated as follows:
\begin{equation}
\Sigma_\star(R) = \Upsilon_\star(R) I_{\rm Ser}(R)= \Upsilon_{\star 0} \tilde{I}_{\rm Ser}(R) \simeq  \Upsilon_{\star 0} \left[ I_{\rm Ser}(R)+ I_{\rm cen}(R)\right], 
 \label{eq:S2RSer}  
\end{equation}
where $I_{\rm cen}(R)$ is the added central component given by Equation~(\ref{eq:IcenR}) to represent the gradient, and $\tilde{I}_{\rm Ser}(R) \simeq I_{\rm Ser}(R)+ I_{\rm cen}(R)$ is the effective distribution as if the ratio were a constant $\Upsilon_{\star 0}$.

\begin{figure}[h] 
\begin{center}
\setlength{\unitlength}{1cm}
\begin{picture}(7,7)(0,0)
\put(-2.5,7.5){\includegraphics{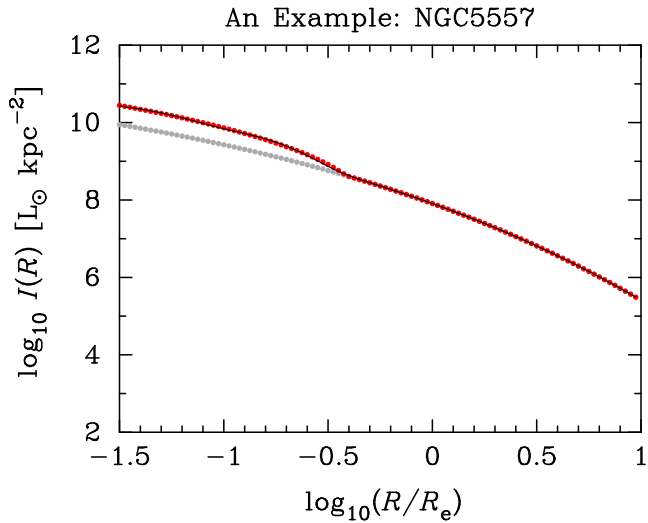}}
\end{picture}
\caption{Example of the S\'{e}rsic profile with $n=4.635$ (gray dots), fitted to the MGE surface brightness distribution of NGC~5557 (Fig.~\ref{SBRMGEser}), and the effective distribution (red dots) defined by multiplying $\Upsilon_\star(R)/\Upsilon_{\star 0}$ (Equation~\ref{eq:MLgrad} with $K=1$) by the S\'{e}rsic function. The black curve fitted to the red dots is the composite distribution that includes an added double-Gaussian model (Equation~\ref{eq:IcenRp}) to accommodate the central radial gradient.}
\label{Sergrad}
\end{center}
\end{figure}

\begin{figure}[h] 
\begin{center}
\setlength{\unitlength}{1cm}
\begin{picture}(9,9)(0,0)
\put(-0.3,-0.8){\includegraphics{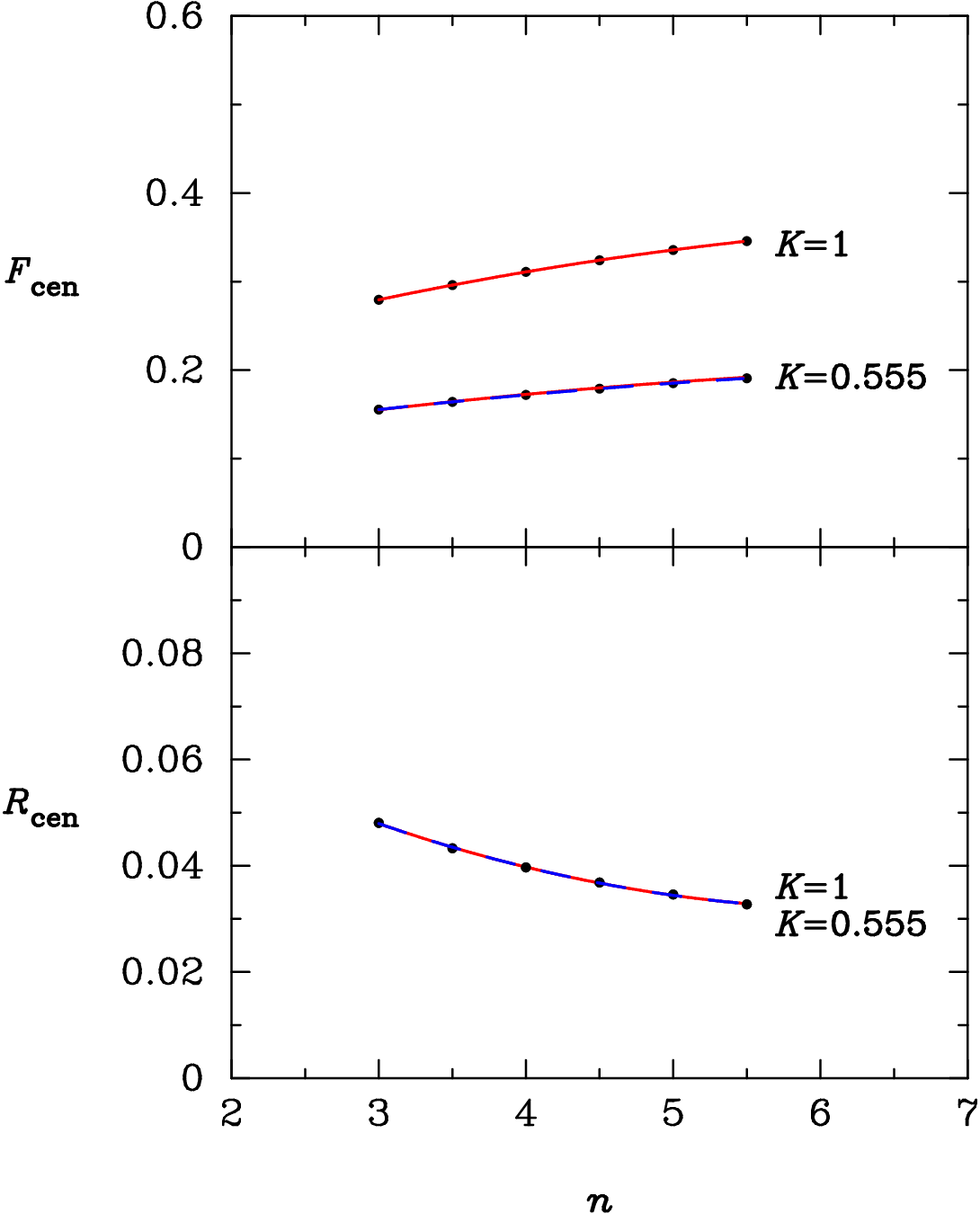}}
\end{picture}
\caption{Relative mass ($F_{\rm cen}$) and the relative core size of the first Gaussian component ($R_{\rm cen}$) of the central double-Gaussian model compared with the main S\'{e}rsic body, numerically derived as a function of S\'{e}rsic index $n$ for gradient strength $K=1$ and $0.555$. At fixed $n$, $F_{\rm cen}(K)=K\times F_{\rm cen}(K=1)$ to a good approximation as shown by the red and blue dashed curves at $K=0.555$ where the blue curve represents the numerical fit results and the red curve is the prediction by $F_{\rm cen}(K=0.555)=0.555\times F_{\rm cen}(K=1)$. At fixed $n$, $R_{\rm cen}$ (in units of $R_{\rm e}$) is independent of $K$ to a good approximation. See the text for further details about the central double-Gaussian component.}
\label{FRcenK}
\end{center}
\end{figure}

For the gradient model under consideration (Equation~\ref{eq:MLgrad}) we can set $I_{\rm cen,2}=0.3I_{\rm cen,1}$ and $\sigma_{\rm cen,2}=3.3\sigma_{\rm cen,1}$ based on numerical investigation. So we rewrite Equation~(\ref{eq:IcenR}) as
\begin{equation}
I_{\rm cen}(R)=I_{\rm cen}\left[ \exp\left(-\frac{R^2}{2R_{\rm cen}^2}\right)+0.3 \exp\left(-\frac{R^2}{2(3.3 R_{\rm cen})^2}\right) \right]   
 \label{eq:IcenRp}  
\end{equation}
with $L_{\rm cen}=8.534 \pi I_{\rm cen} R_{\rm cen}^2$, which now has just two free parameters $R_{\rm cen}$ and $L_{\rm cen}$ (or $I_{\rm cen}$). Let $F_{\rm cen}\equiv L_{\rm cen}/L_{\rm Ser}$ where $L_{\rm Ser}$ is the total luminosity of the S\'{e}rsic light distribution.  Then, we can solve numerically Equation~(\ref{eq:S2RSer}) to find the best-fit $F_{\rm cen}$ and $R_{\rm cen}$ for a given S\'{e}rsic profile. Figure~\ref{Sergrad} shows an example with gradient strength $K=1$ for a S\'{e}rsic index $n=4.635$.
From numerical results at various values of $n$ we find
\begin{equation}
F_{\rm cen}(n)=0.31091 + 0.02813(n-4) -0.00331(n-4)^2,  
 \label{eq:Fcen}  
\end{equation}
and 
\begin{equation}
R_{\rm cen}(n)=0.03976  -0.00675(n-4)+ 0.00143(n-4)^2     
 \label{eq:Rcen}  
\end{equation}
in units of $R_{\rm e}$ for the range $3\le n \le 5.5$ with $K=1$. With other values of $K$, $R_{\rm cen}(n)$ remains unchanged and $F_{\rm cen}(n)|_K=K\times F_{\rm cen}(n)|_{K=1} $, to a good approximation. Figure~\ref{FRcenK} shows numerical examples fitted with these curves. 

The total stellar mass for the distribution of Equation~(\ref{eq:S2RSer}) is
\begin{equation}
M_\star = \Upsilon_{\star 0}(L_{\rm Ser}+L_{\rm cen}) = M_{\rm \star Ser}+M_{\rm \star cen} = (1+F_{\rm cen})M_{\rm \star Ser},                             
 \label{eq:Mstartot}  
\end{equation}
where we define $M_{\rm \star Ser}\equiv \Upsilon_{\star 0}L_{\rm Ser}$ and $M_{\rm \star cen}\equiv \Upsilon_{\star 0} L_{\rm cen}$. The projected mass within the half-light radius $R_{\rm e}$ is then given by
\begin{equation}
M_{\rm \star e}= \frac{1}{2} M_{\rm \star Ser} + M_{\rm \star cen} = \frac{1 + 2 F_{\rm cen}}{2(1 + F_{\rm cen})} M_\star,                               
 \label{eq:Mstare}  
\end{equation}
which is $>0.5 M_\star$ for $F_{\rm cen}>0$. The projected stellar mass density within $R_{\rm e}$, $\mu_{\rm \star e}\equiv M_{\rm \star e}/(\pi R_{\rm e}^2)$, is related to $M_\star$ as follows:
\begin{equation}
\log_{10}M_\star = \log_{10}\mu_{\rm \star e} + \log_{10}(2\pi R_{\rm e}^2) - \log_{10}\left(\frac{1+2F_{\rm cen}}{1 + F_{\rm cen}}\right).                     
 \label{eq:mustare}  
\end{equation}

\begin{figure}[h] 
\begin{center}
\setlength{\unitlength}{1cm}
\begin{picture}(7,7)(0,0)
\put(-2.5,7.5){\includegraphics{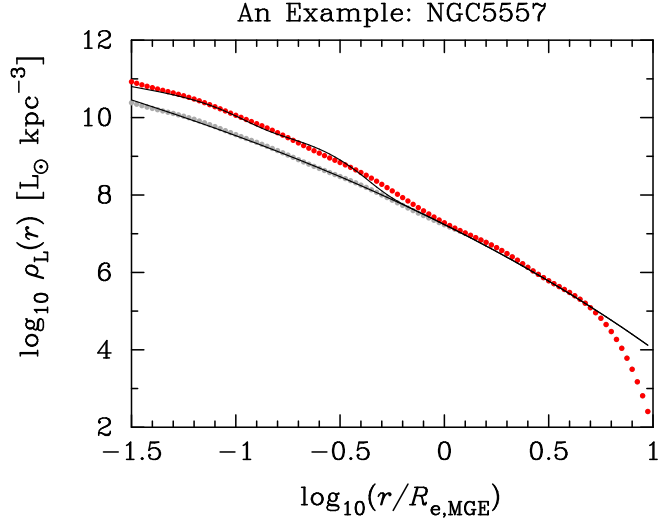}}
\end{picture}
\caption{Example of the volume light distributions obtained by deprojecting the observed and the effective (i.e.\ accommodating $M_\star/L$ gradient in the central region) surface brightness distributions through two approaches. The gray and red dots represent the results through the MGE model, i.e.\ the deprojection of Figure~\ref{SBRMGE}, while the black curves represent the results through the S\'{e}rsic model with the added double-Gaussian model, i.e.\ the deprojection of Figure~\ref{Sergrad}.}
\label{rhoLgrad}
\end{center}
\end{figure}

For a SDSS pure-bulge galaxy with its measured $n$, $R_{\rm e}$ and $\sigma_{\rm e}$ we can estimate $M_\star$ (or $\Upsilon_{\star 0}$) and mass densities using the FMP derived for the \atl galaxies as a function of $K$. We first estimate $\mu_{\rm \star e}$ using the FMP as a function of $K$ (see \S 3.2). We estimate $F_{\rm cen}(n)$ as a function of $K$ and $R_{\rm cen}(n)$ independent of $K$ as described above. With the set \{$\mu_{\rm \star e}$, $F_{\rm cen}(n)$, $R_{\rm cen}(n)$\} we can estimate $M_\star$ using Equation~(\ref{eq:mustare}) and then the projected density (Equation~\ref{eq:S2RSer}), the volume density and the stellar mass within a spherical radius with Equations~(\ref{eq:IcenR})---(\ref{eq:Lcenr}) for the central component and with the corresponding equations for the S\'{e}rsic body found in the literature (e.g., Appendix~A of \cite{Chae12}). Figure~\ref{rhoLgrad} shows an example of the deprojected 3-dimensional light distributions using NGC~5557. It shows the deprojected profiles of the observed surface brightness distribution and the effective distribution accommodating a central $M_\star/L$ gradient using two approaches, i.e.\ the MGE model with effective coefficients (\S 2.7.1) and the S\'{e}rsic plus central double-Gaussian model considered in this section. These two approaches give overall consistent results. Some minor discrepancies between the two, particularly near $0.4 R_{\rm e}$ have no impact on this work because the current large uncertainties in the $M_\star/L$ gradient do not warrant a very precise model.

\section{Modeling Procedures and Results}

Our goal is to construct Monte Carlo sets of models for baryonic and DM mass distributions under the \LCDM paradigm, or for baryonic mass distribution plus MOND IF under the MOND paradigm. We derive a Monte Carlo set of models for each galaxy using the measured light distribution, a range of $M_\star/L$ gradients, and the measured VD profile for the \atl galaxies, or the aperture VD for the SDSS galaxies. We consider the \atl galaxies first and then the SDSS galaxies. In modeling the SDSS galaxies we use an FMP relation based on the modeling results for the \atl galaxies.

\subsection{For the \atl Galaxies}

We use the VDP constructed from the measured LOSVD map as described in \S 2.7.2. Each radial bin at radius $R_i$ contains at least 21 LOSVDs from which we obtain the statistically weighted mean (or median) $\sigma_{\rm los}^{\rm obs}(R_i)$ and the standard uncertainty $s_i$. We fit $\Upsilon_\star(R)$ and the DM mass distribution $M_{\rm DM}(r)$, or MOND IF, by minimizing the `goodness-of-fit' statistic chi-squared
\begin{equation}
 \chi^2 = \sum_{i=1}^{N_{\rm bin}} \frac{\left[\sigma_{\rm los}^{\rm obs}(R_i)-\sigma_{\rm los}(R=R_i)\right]^2}{s_i^2},
 \label{eq:chisq}
\end{equation}
where $\sigma_{\rm los}(R=R_i)$ is the LOSVD (Equation~\ref{eq:losvd}) predicted by the model under consideration. 

To obtain a Monte Carlo set of models for each galaxy under the \LCDM paradigm, we first draw DM halo parameters \{$M_{200}$,~$c_{\rm NFW}$,~$\alpha$\}. For halo mass $M_{200}$ we take a Gaussian deviate of $\log_{10}M_{200}$ using its empirical correlation with a fixed IMF-based stellar mass with a typical scatter of $\sim 0.2$~dex (\S 2.5, Figure~\ref{SMDM}). For $c_{\rm NFW}$ we take a Gaussian deviate using $c_{\rm NFW}=(7.192/(1+z))(M_{200}/(10^{14}{\rm M}_{\odot}/h))^{-\varepsilon}$, where $z$ is the redshift of the galaxy, $h=0.7$ and $\varepsilon=0.114$, with an rms scatter of $0.15$. For the profile slope $\alpha$ we take a uniform deviate from the range $0<\alpha<1.8$. Note that $c_{200}$ follows from $\alpha$ and $c_{\rm NFW}$ using the weak-lensing constraint for $r > 0.2 r_{200}$ (see \S 2.5). We also take a $M_\star/L$ gradient strength $K$ from $0\le K < 1.5$. For the MOND case we draw $\nu$ (Equation~\ref{eq:IFnu}) from $0<\nu \le 2$ and $a_0$ from $0.5 < a_0/(10^{-10}\rm{m}~\rm{s}^{-2}) <1.9$. For each random draw of $\{M_{200},\hspace{1ex}c_{\rm NFW},\hspace{1ex}\alpha,\hspace{1ex} K\}$ or $\{\nu,\hspace{1ex}a_0,\hspace{1ex}K\}$, $\Upsilon_{\star 0}$ and the given anisotropy model are varied to minimize $\chi^2$. Anisotropy values are allowed to vary between $-2$ and $0.7$ so that the ratio $\sigma_{\rm t}^2/\sigma_{\rm r}^2$ varies between $1/3$ and $3$.

\begin{figure}[h] 
\begin{center}
\setlength{\unitlength}{1cm}
\begin{picture}(7,7)(0,0)
\put(-2.5,7.5){\includegraphics{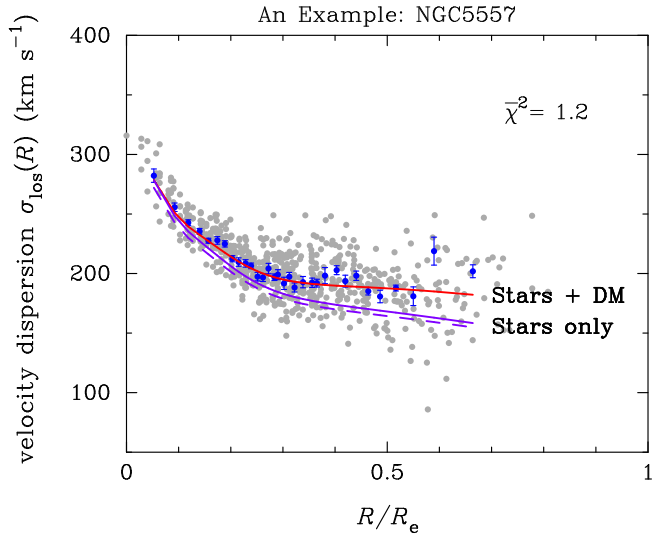}}
\end{picture}
\caption{Example of the observed LOSVDs and their successful modeling under the $\Lambda$CDM paradigm. The gray dots are individual measurements on the sky and blue dots represent binned, error-weighted means. The red curve is the prediction by the successful model which has a reduced chi-squared of $\bar{\chi}^2 =1.2$. The purple dashed curve is the prediction by the stellar mass distribution only. The purple solid curve is the readjusted prediction by the stellar mass distribution so that the innermost VDs are reproduced.}
\label{VDP1}
\end{center}
\end{figure}

For a good model we expect the reduced chi-squared $\bar{\chi}^2=\chi^2/N_{\rm dof}\sim 1$ where the degree of freedom $N_{\rm dof}=N_{\rm bin}-N_{\rm free}\approx N_{\rm bin}$ which is typically 20--30. With constant anisotropies, good fits ($\bar{\chi}^2\lesssim 2$) cannot be found for about 40\% of galaxies. With varying gOM anisotropies (Equation~\ref{eq:gOM}) good fits ($\bar{\chi}^2\lesssim 2$) are found for most galaxies.  Figure~\ref{VDP1} shows a detailed example of successful models with $\bar{\chi}^2 \sim 1$ overplotted on the observed VDP of NGC~5557 for the \LCDM case. Figure~\ref{VDP1} also illustrates the unsuccessful prediction without DM. Figure~\ref{VDP_A3D} exhibits the formal best-fit (i.e.\ $\bar{\chi}^2=\bar{\chi}_{\rm min}^2$) models for all 24 \atl pure-bulges for the \LCDM case. Very similar results are obtained for the MOND case as well.

For each galaxy, 800 Monte Carlo models are produced within the prior parameter ranges. But we keep only those models with $\bar{\chi}^2 < 2 \bar{\chi}^2_{\rm min}$ where $\bar{\chi}^2_{\rm min}$ is the minimum value obtained for the best-fit model. Using these models we calculate the likely range of $M_\star/L$ gradient strength $K$, stellar masses, and the FMP, which are described below. In companion and subsequent papers we use the sets to investigate the likely ranges of mass profiles, DM fractions, MOND IFs, VD anisotropies, and the radial acceleration relation.

\begin{figure} 
\begin{center}
\setlength{\unitlength}{1cm}
\begin{picture}(10,16)(0,0)
\put(-2.3,-1.){\includegraphics{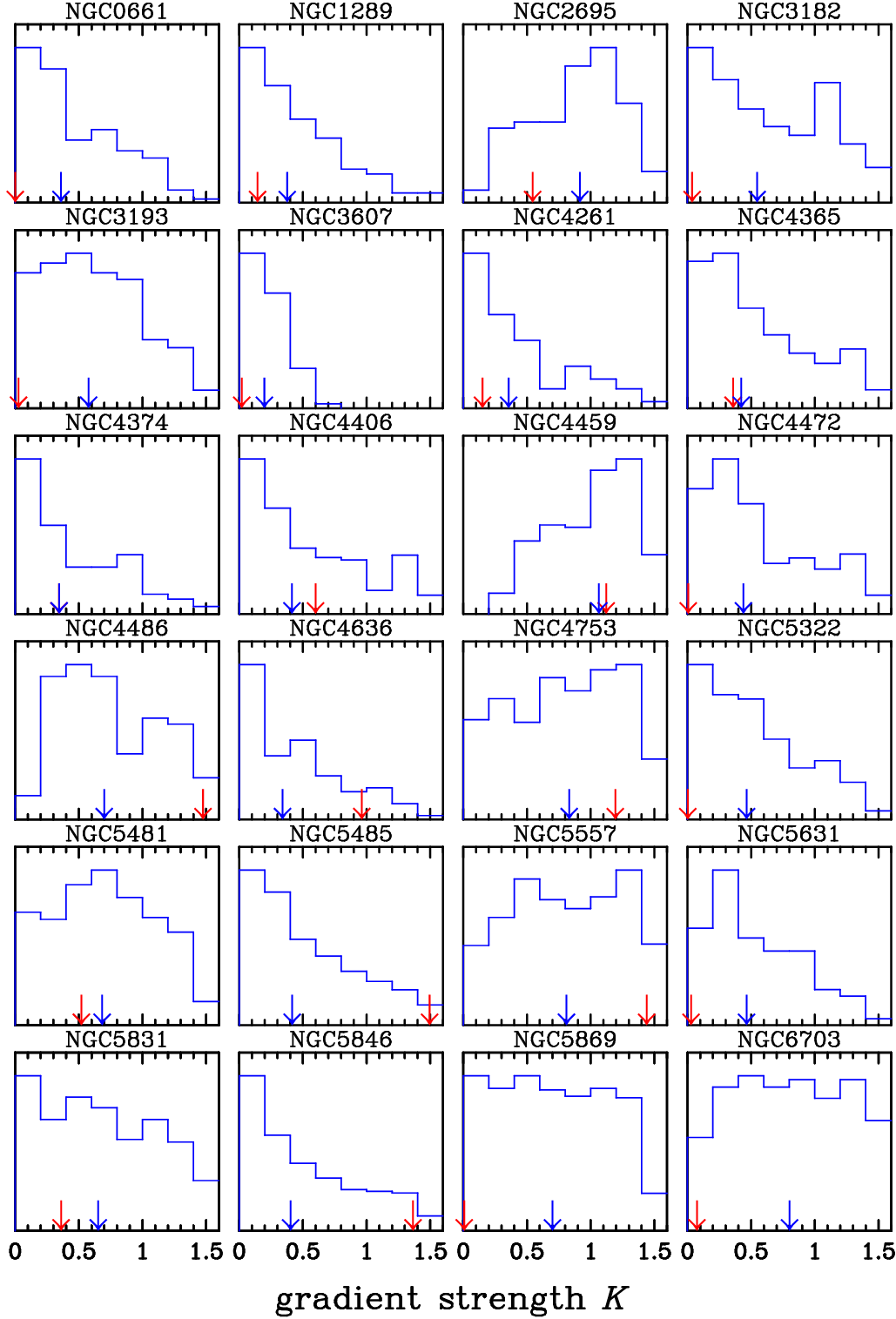}}
\end{picture}
\caption{Histograms showing distributions of $M_\star/L$ gradient strength $K$ (Equation~\ref{eq:MLgrad}) from the Monte Carlo sets of models satisfying $\bar{\chi}^2 < 2 \bar{\chi}^2_{\rm min}$ for the \atl galaxies shown in Fig.~\ref{VDP_A3D}. The red arrows indicate best-fit values (i.e.\ for the best-fit models satisfying $\bar{\chi}^2=\bar{\chi}^2_{\rm min}$), while the blue arrows indicate median values from the models.}
\label{KhistAll}
\end{center}
\end{figure}

\begin{figure} 
\begin{center}
\setlength{\unitlength}{1cm}
\begin{picture}(12,12)(0,0)
\put(-2.,10.){\includegraphics{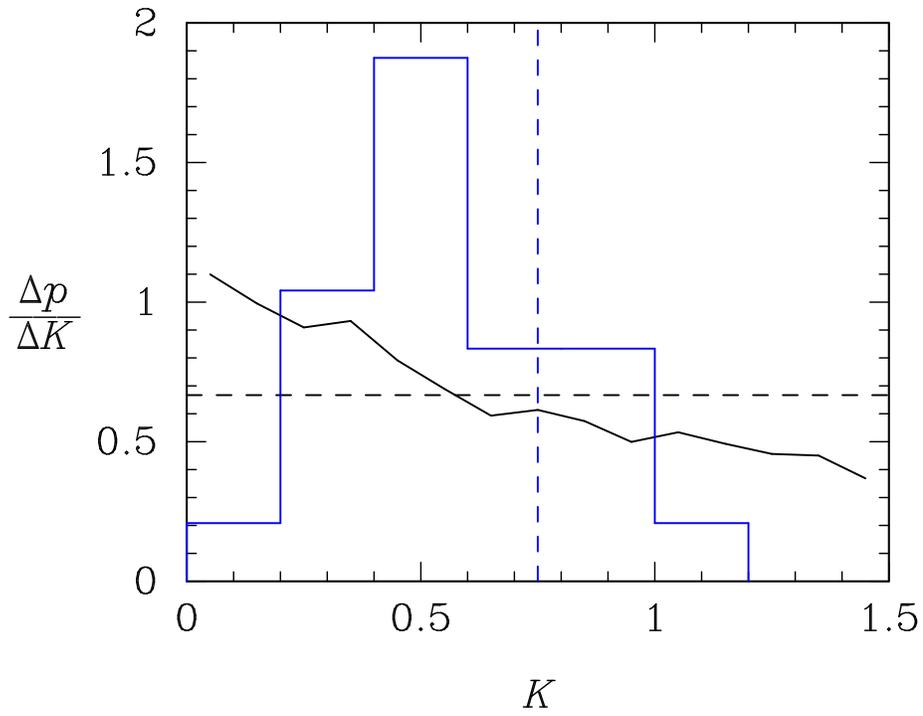}}
\end{picture}
\caption{Probability distribution of $K$ values for the 24 \atl galaxies shown in Figure~\ref{KhistAll}. The black solid curve is the averaged distribution of the individual distributions, while the horizontal dashed line is the expected null result. The blue histogram represents the distribution of the individual medians, while the vertical dashed line is the expected null result. These results show that a strong gradient ($K\gtrsim 1$) is less likely.}
\label{Kmed}
\end{center}
\end{figure}

\begin{figure} 
\begin{center}
\setlength{\unitlength}{1cm}
\begin{picture}(10,16)(0,0)
\put(-2.3,-1.){\includegraphics{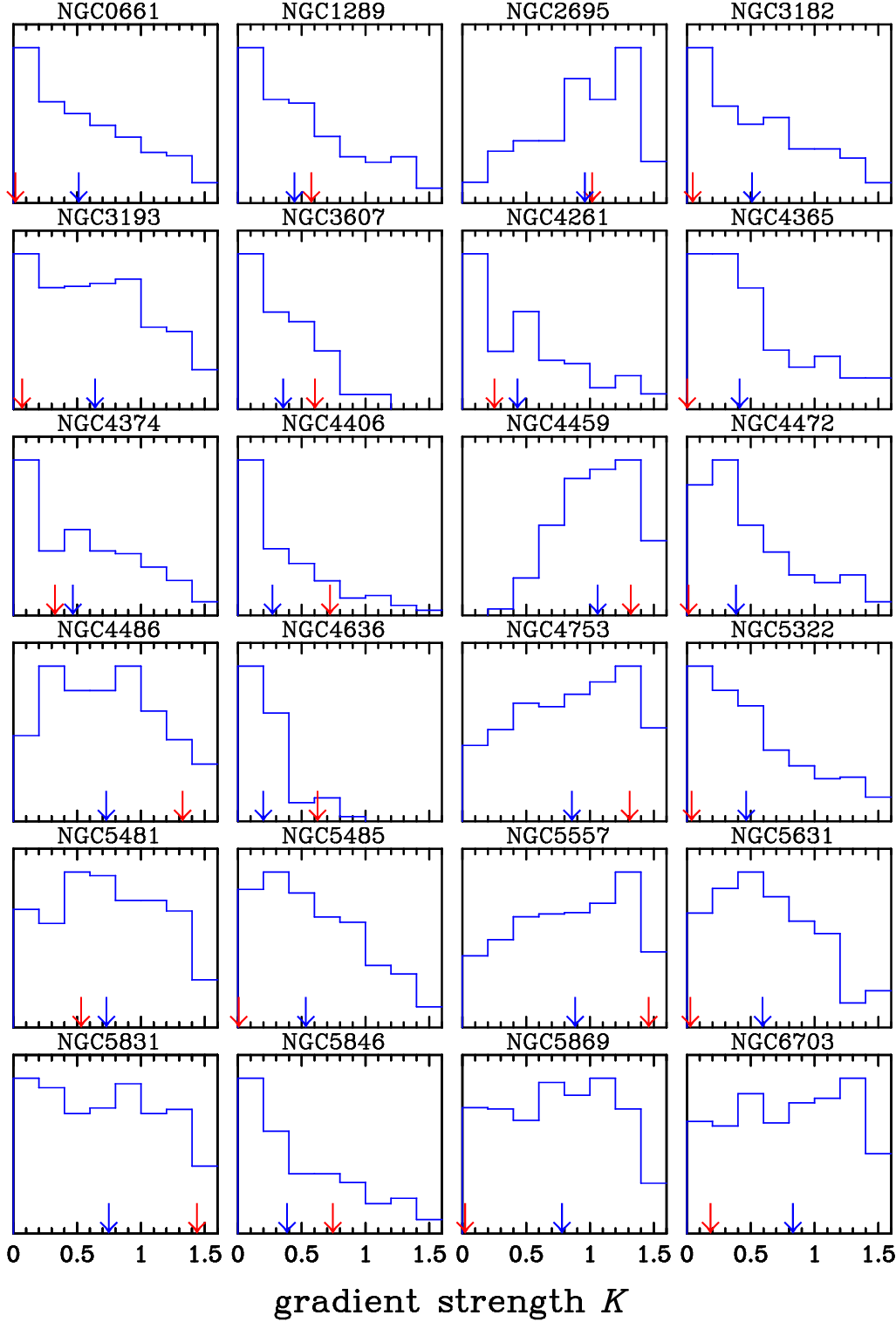}}
\end{picture}
\caption{Same as Figure~\ref{KhistAll} but under the MOND paradigm.}
\label{KhistMDAll}
\end{center}
\end{figure}

Figure~\ref{KhistAll} shows the distributions of $K$ from the models satisfying $\bar{\chi}^2 < 2 \bar{\chi}^2_{\rm min}$ for all 24 galaxies. There are significant galaxy-to-galaxy scatters for $K$. There are also severe degeneracies of $K$ in many cases (e.g.\ NGC~5557, NGC~5869, NGC~6703 in particular). Due to these degeneracies and numerical limitations, best-fit and median values of $K$ are sometimes discrepant. Considering the degeneracies the median values are more likely to be meaningful. Figure~\ref{Kmed} exhibits the distribution of the medians and the averaged probability density of $K$ based on the individual distributions of Figure~\ref{KhistAll}. The distribution of the median values is peaked near $\langle K \rangle = 0.5$. We estimate $\langle K \rangle = 0.53^{+0.05}_{-0.04}$ along with an rms scatter of $s_K = 0.17^{+0.06}_{-0.03}$ based on a bootstrap method using the distribution. This value is lower than the null result $0.75$ (recall that our prior range is $0\le K <1.5$). The distribution of the averaged probability density is tilted from the horizontal line showing that a strong gradient ($K \gtrsim 1$) is less likely. This result agrees well with the intermediate strength $K=0.555$ suggested by \cite{Ber18}, which is weaker than the \cite{vD17} strength $K=1$. Very similar results are obtained under the MOND paradigm, as shown in Figure~\ref{KhistMDAll}.

\subsection{The FMP Relation with a Stellar Mass-to-light Ratio Gradient}

\begin{figure}[h] 
\begin{center}
\setlength{\unitlength}{1cm}
\begin{picture}(8,8)(0,0)
\put(-0.8,-2.8){\includegraphics{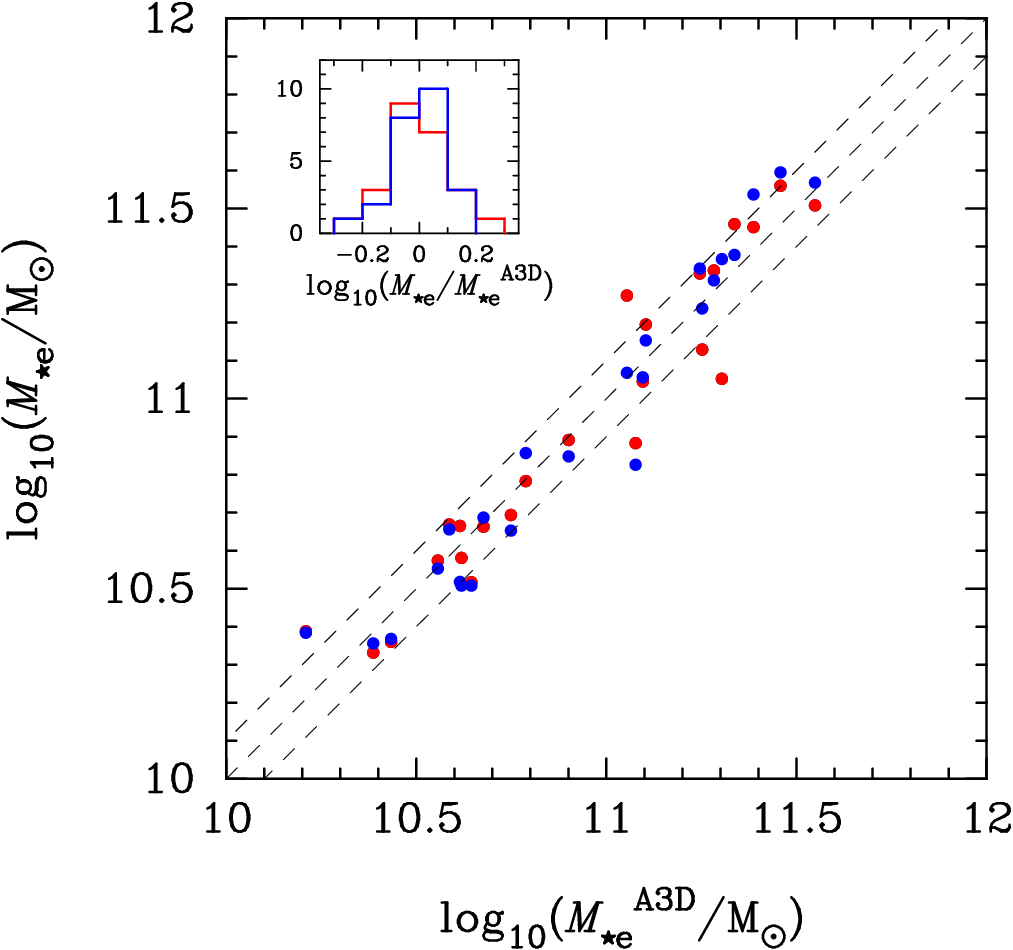}}
\end{picture}
\caption{Consistency of the projected stellar mass within $R_{\rm e}$ between the ATLAS$^{\rm 3D}$ reported values and our modeling results for the 24 pure-bulge galaxies under the \LCDM paradigm, with the assumption of constant $M_\star/L$. The red points represent the results for the best-fit models ($\bar{\chi}^2 = \bar{\chi}^2_{\rm min}$), while the blue points represent the median results for the models satisfying $\bar{\chi}^2 < 2 \bar{\chi}^2_{\rm min}$.}
\label{Me2}
\end{center}
\end{figure}

Elliptical galaxies are supposed to be in dynamical equilibrium. According to the virial theorem \citep{BT} we expect correlations between a fiducial radius (which is chosen to be $R_{\rm e}$ here), dynamical total mass ($M_{\rm dyn,e}$), and an observable mean VD squared ($\sigma_{\rm e}^2$ is a proxy for mean kinetic energy), both within that radius: $M_{\rm dyn,e}/R_{\rm e} \propto \sigma_{\rm e}^2$. Let $M_{\rm \star e}$ be the projected stellar mass within $R_{\rm e}$. Then we expect a good correlation between $M_{\rm \star e}$ and $M_{\rm dyn,e}$, and hence we expect the following correlation
\begin{equation}
  w = a + b(x -x_0) + c(y -y_0),
  \label{eq:FMP}
\end{equation}
where $x \equiv \log_{10}\sigma_e$, $y \equiv \log_{10}R_{\rm e}$ and $w \equiv \log_{10} \mu_{\rm \star e}=\log_{10}\left[M_{\rm \star e}/(\pi R_{\rm e}^2)\right]$. Equation~(\ref{eq:FMP}) is a representation of the fundamental plane of early-type galaxies \citep{DD,Dre}.  (Here $\sigma_{\rm e}$ is in km~s$^{-1}$, $R_{\rm e}$ is in kpc, and mass density is in M$_\odot$~kpc$^{-2}$.) We take $R_{\rm e}$ and $\sigma_e$ (the aperture VD within $R_{\rm e}$) from \cite{Cap13a}. For a case with a constant stellar mass-to-light ratio (denoted by $\Upsilon_{\star 0}$), we estimate $M_{\star{\rm e}}$ using an empirical relation $M_{\star{\rm e}}\approx 0.594 M_{\star{\rm MGE}}$ derived for the 24 pure-bulge galaxies, where $M_{\star{\rm MGE}}$ is the analytic stellar mass for the MGE distribution, i.e.\  $M_{\star{\rm MGE}}= \Upsilon_{\star 0} L_{\rm MGE}$. Here, we can take $\Upsilon_{\star 0}$ and $L_{\rm MGE}$ from the \atl reported results \citep{Cap13a,Cap13b}. We can also estimate $M_{\star{\rm e}}$ using our Monte Carlo sets, which are based on radially varying anisotropies (Equation~\ref{eq:gOM}). Note that the \atl modeling was based on radially constant anisotropies, although they used oblate spheroids rather than spherical models. The two results agree, as shown in Figure~\ref{Me2}, although there are individual scatters.

Our derivation of the parameters of the plane is based on Equation~(7) of \cite{Cap13a}. Namely, we minimize the quantity
\begin{equation}
\Delta^2 = \sum_{j=1}^{N_{\rm gal}} \frac{\left[a+b(x_j-x_0)+c(y_j-y_0)-w_j\right]^2}
      {(b s_{x_j})^2 + (c s_{y_j})^2 + ( s_{w_j})^2 + \varepsilon_w^2 },
 \label{eq:Delsq}  
\end{equation}
 where $x_0=2.11$ and $y_0=0.301$, as in \cite{Cap13a}, $s_{x_j}$ and $s_{y_j}$ are the \atl reported errors and $s_{w_j}$ is estimated using the standard error propagation, assuming normal errors. In Equation~(\ref{eq:Delsq}), $\varepsilon_w$ is the intrinsic scatter of $w$ that is estimated so that the minimized $\Delta^2$ is equal to $N_{\rm gal}-4$ (degree of freedom). We estimate the errors of the parameters $a$, $b$, and $c$ by generating Monte Carlo sets of data $\mathscr{D}^{(i)}=$\{$x_j^{(i)}$,~$y_j^{(i)}$,~$w_j^{(i)}$\} ($j=1,...,N_{\rm gal}$) and obtaining \{$a^{(i)}$,~$b^{(i)}$,~$c^{(i)}$\} fitted to $\mathscr{D}^{(i)}$ ($i=1,...,900$).

\begin{figure}[h] 
\begin{center}
\setlength{\unitlength}{1cm}
\begin{picture}(8,8)(0,0)
\put(-1.,-2.6){\includegraphics{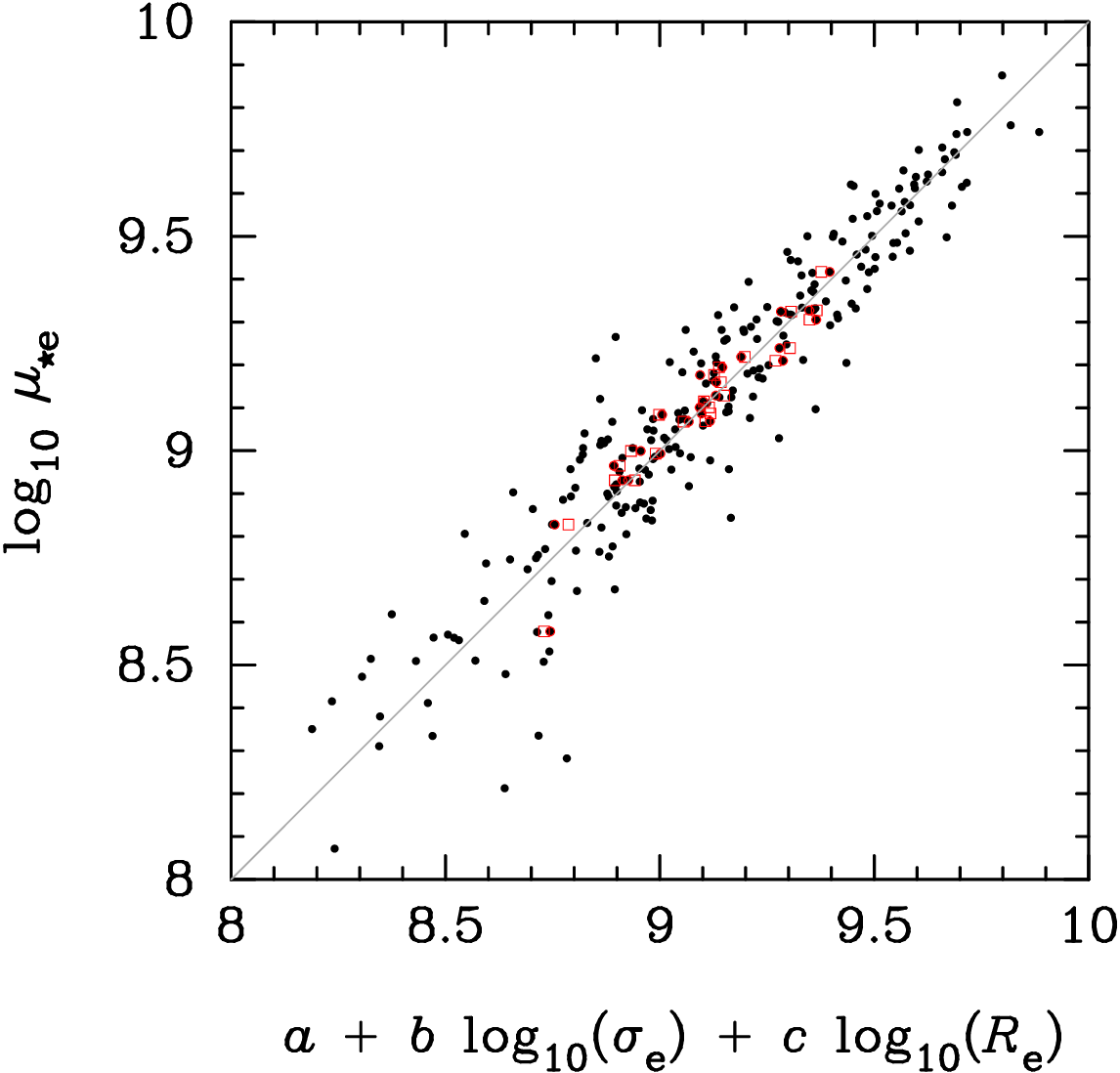}}
\end{picture}
\caption{Fundamental mass plane (FMP) relation for \atl galaxies among $\log_{10}(\sigma_{\rm e})$, $\log_{10}(R_{\rm e})$ and $\log_{10}\mu_{\rm \star e} \equiv \log_{10}[M_{\rm \star e}/(\pi R_{\rm e}^2)]$. Here, $\sigma_{\rm e}$ is in km~s$^{-1}$, $R_{\rm e}$ is in kpc, and $M_{\rm \star e}$ is in M$_\odot$ (solar masses). For the zero-points of $\log_{10}(\sigma_{\rm e})$ and $\log_{10}(R_{\rm e})$ we take $x_0=2.11$ and $y_0=0.301$. The black dots represent 258 ATLAS$^{\rm 3D}$ early-type galaxies. The black dots within red circles are 24 pure-bulge galaxies. The open red squares represent the FMP derived only for the pure-bulges. }
\label{FMP}
\end{center}
\end{figure}

 For the case of constant $M_\star/L$ [$K=0$ in Equation~(\ref{eq:MLgrad})] the FMP parameter values obtained for all \atl galaxies (except for two galaxies for which not all necessary galaxy parameter values are provided) are $a=9.174\pm 0.007$, $b=2.22\pm 0.05$, $c=-1.13\pm 0.04$ (with $\epsilon_w=0.088$), and an rms scatter of $\Delta_w=0.145$.  For the 24 pure-bulge galaxies we have $a=9.141\pm 0.058$, $b=2.27 \pm 0.34$, $c=-1.05\pm 0.21$ (with $\varepsilon_w = 0$), and $\Delta_w=0.062$. These two results are consistent with each other and the virial expectation, i.e.\ $b=2$ and $c=-1$. Figure~\ref{FMP} shows this explicitly: the two FMP estimates of $w_j$ given the measured $x_j$ and $y_j$ are very similar. Our estimate of the FMP is also similar to that reported in \cite{Cap13a}, based on a dynamical mass $M_{\rm JAM}$ that is obtained by multiplying their dynamical (stellar plus dark) mass-to-light ratio by their total luminosity; their reported parameters in their Figure~12 (second panel) can be translated to $a=9.197\pm 0.004$, $b=1.928\pm 0.026$, and $c=-1.036\pm 0.018$ if we use $M_{\rm JAM}/(2\pi R_{\rm e}^2)$ as a proxy for $\mu_{\rm \star e}$.

\begin{figure}[h] 
\begin{center}
\setlength{\unitlength}{1cm}
\begin{picture}(7,7)(0,0)
\put(-2.5,7.5){\includegraphics{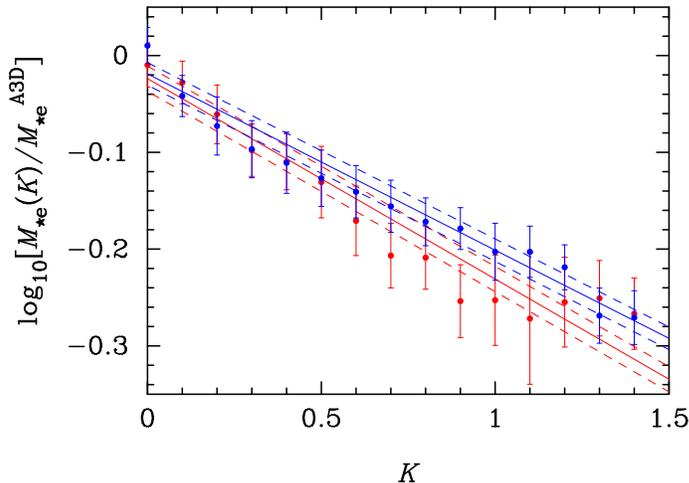}}
\end{picture}
\caption{Dependence of the dynamical estimate of the projected stellar mass within $R_{\rm e}$ ($M_{\rm \star e}$) on $M_\star/L$ gradient strength ($K$) in the 24 pure-bulge \atl galaxies.  Here, $M_{\rm \star e}$ is the estimate as a function of $K$ based on our Jeans modeling results under the \LCDM paradigm, whereas $M_{\rm \star e}^{\rm A3D}$ is the estimate based on the \atl JAM modeling results for $K=0$. The red and blue points represent the results, respectively, for the best-fit and the median models. The solid lines are the least-square fit results. The dashed lines represent the uncertainties of the y-intercept. The fit results are as follows: $y=a'+b' K$ with $a'=-0.024\pm 0.013$ and $b'=-0.21 \pm 0.02$ (red, best-fit), or $a'=-0.019\pm 0.012$ and $b'=-0.18 \pm 0.02$ (blue, median).}
\label{MeK_A3D}
\end{center}
\end{figure}

\begin{figure}[h] 
\begin{center}
\setlength{\unitlength}{1cm}
\begin{picture}(7,7)(0,0)
\put(-2.5,7.5){\includegraphics{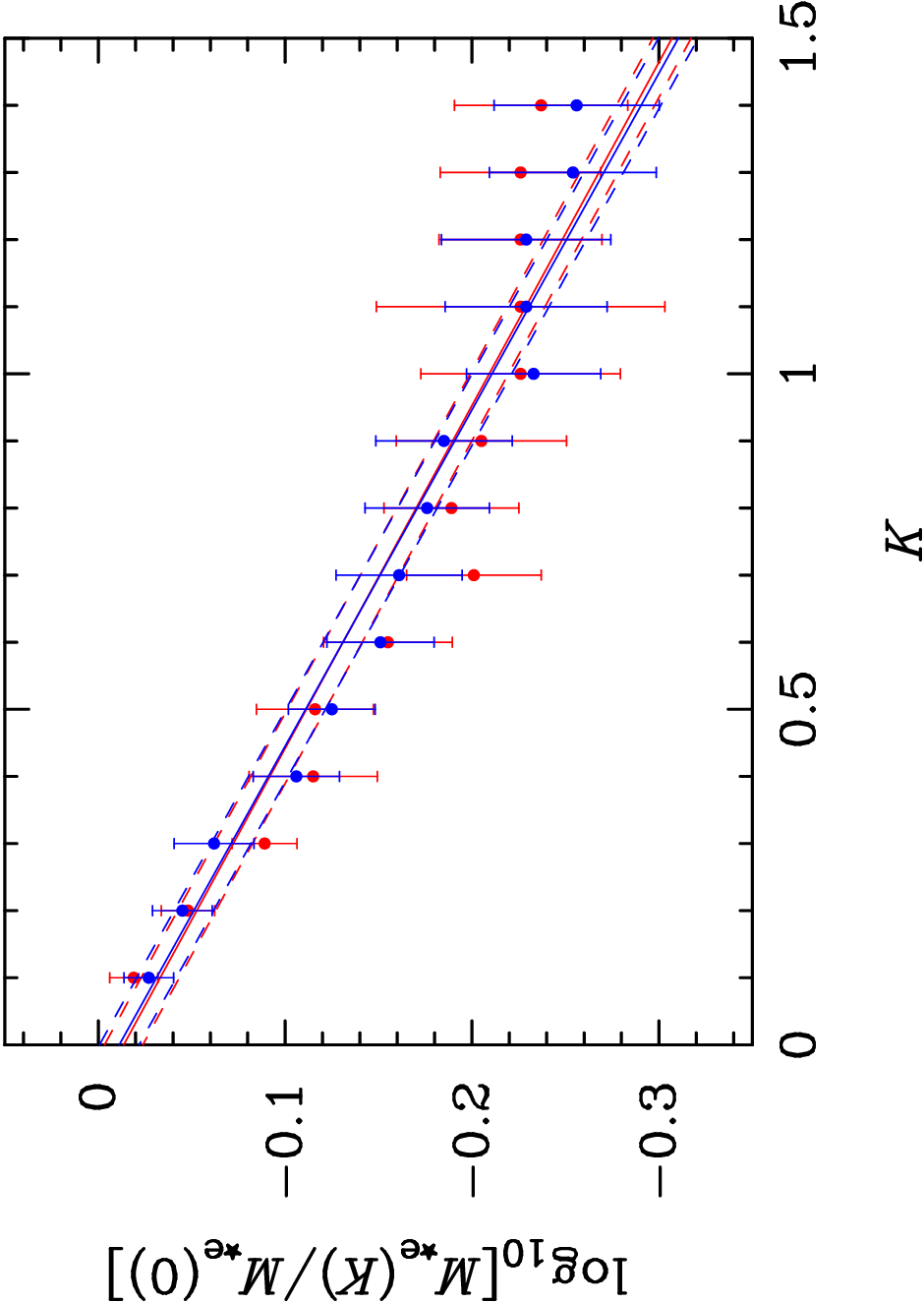}}
\end{picture}
\caption{Same as Figure~\ref{MeK_A3D}, but with respect to our own estimate of $M_{\rm \star e}$ at $K=0$. The fit results are as follows: $y=a'+b' K$ with $a'=-0.014\pm 0.010$ and $b'=-0.20 \pm 0.02$ (red, best-fit), or $a'=-0.011\pm 0.011$ and $b'=-0.20 \pm 0.02$ (blue, median).}
\label{MeK}
\end{center}
\end{figure}

\begin{figure}[h] 
\begin{center}
\setlength{\unitlength}{1cm}
\begin{picture}(7,7)(0,0)
\put(-2.5,7.5){\includegraphics{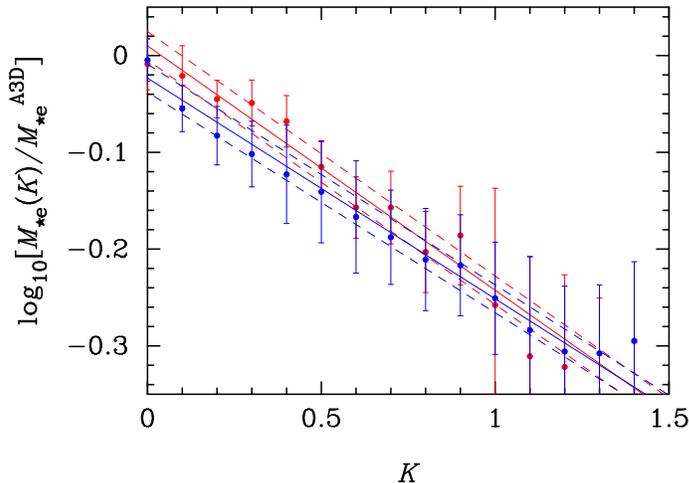}}
\end{picture}
\caption{Same as Figure~\ref{MeK_A3D} but under the MOND paradigm. The fit results are as follows: $y=a'+b' K$ with $a'=0.010\pm 0.015$ and $b'=-0.25 \pm 0.03$ (red, best-fit), or $a'=-0.023\pm 0.014$ and $b'=-0.23 \pm 0.03$ (blue, median).}
\label{MeMDK_A3D}
\end{center}
\end{figure}

\begin{figure}[h] 
\begin{center}
\setlength{\unitlength}{1cm}
\begin{picture}(7,7)(0,0)
\put(-2.5,7.5){\includegraphics{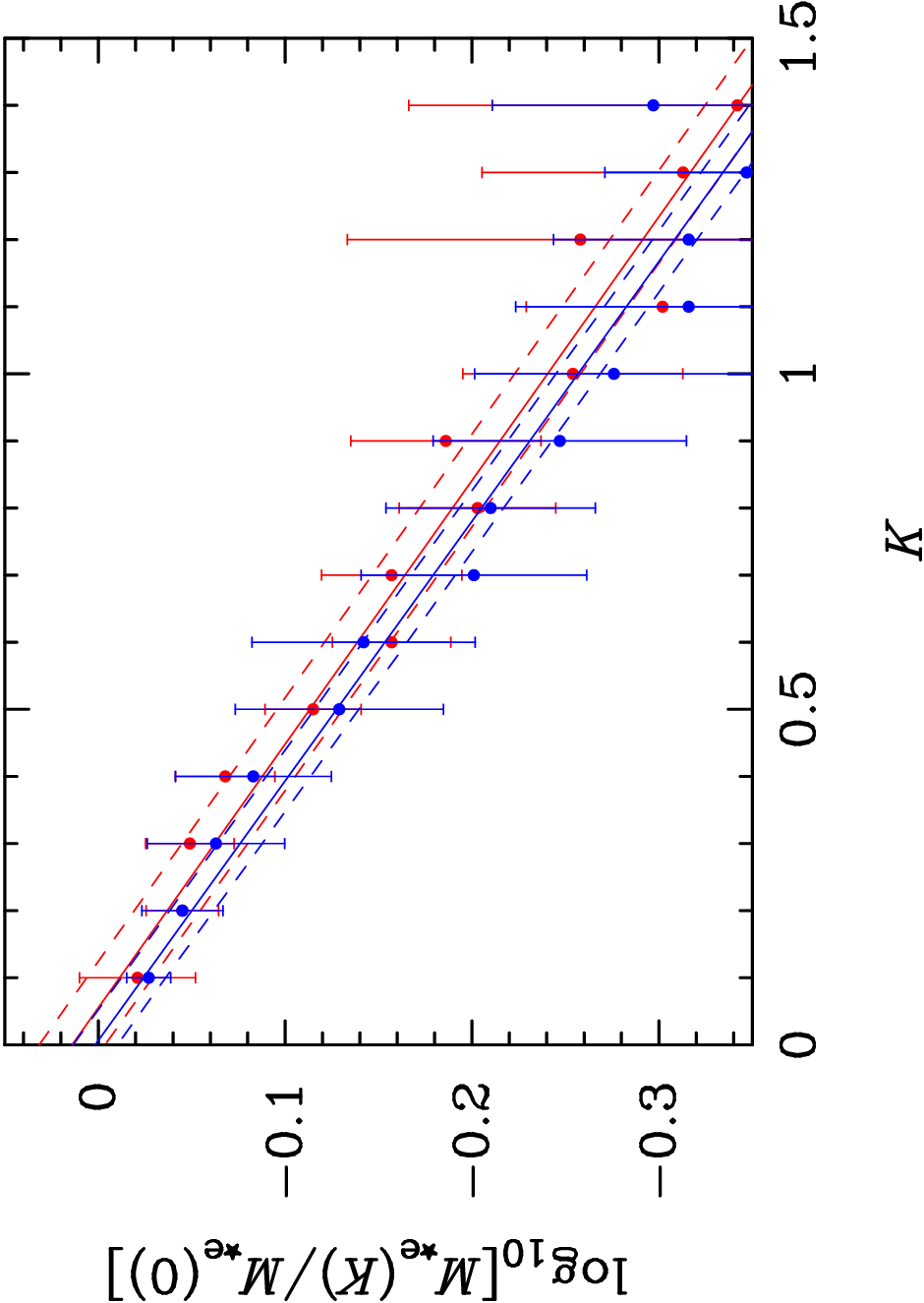}}
\end{picture}
\caption{Same as Figure~\ref{MeMDK_A3D} but with respect to our own estimate of $M_{\rm \star e}$ at $K=0$. The fit results are as follows: $y=a'+b' K$ with $a'=0.014\pm 0.018$ and $b'=-0.25 \pm 0.03$ (red, best-fit), or $a'=0.002\pm 0.012$ and $b'=-0.26 \pm 0.03$ (blue, median).}
\label{MeMDK}
\end{center}
\end{figure}

 The pure-bulge FMP allows an accurate estimate of $M_{\star{\rm e}}$ for pure-bulge galaxies with measured $\sigma_{\rm e}$ and $R_{\rm e}$ under the assumption of constant $M_\star/L$. If there is a $M_\star/L$ gradient ($K\ne 0$), we expect a systematic shift of $M_{\star{\rm e}}$ as demonstrated in \cite{Ber18}. We estimate the systematic shift with $K$ using our Monte Carlo sets with $K\ne 0$ as shown in Figure~\ref{MeK_A3D}. We find a linear relation $\log_{10} \left[M_{\star{\rm e}}(K)/M_{\star{\rm e}}^{\rm A3D} \right] = a' + b' K$ with $a'=-0.024\pm 0.013$ and $b'=-0.21\pm 0.02$ (for the best-fit models), or $a'=-0.019\pm 0.012$ and $b'=-0.18\pm 0.02$ (for the median models), under the \LCDM paradigm, where $M_{\star{\rm e}}^{\rm A3D}$ is from the \atl \citep{Cap13a,Cap13b} modeling results. We obtain similar results (Figure~\ref{MeK}) with respect to our own estimate of the mass at $K=0$ as follows: $\log_{10} \left[M_{\star{\rm e}}(K)/M_{\star{\rm e}}(K=0) \right] = a' + b' K$ with $a'=-0.014\pm 0.010$ and $b'=-0.20\pm 0.02$ (for the best-fit models), or $a'=-0.011\pm 0.011$ and $b'=-0.20\pm 0.02$ (for the median models). Under the MOND paradigm the corresponding results are as follows, as shown in Figures~\ref{MeMDK_A3D} and \ref{MeMDK}: $a'=0.010\pm 0.015$ and $b'=-0.25\pm 0.03$ (for the best-fit models), or $a'=-0.023\pm 0.014$ and $b'=-0.23\pm 0.03$ (for the median models), with respect to $M_{\star{\rm e}}^{\rm A3D}$; and, $a'=0.014\pm 0.018$ and $b'=-0.25\pm 0.03$ (for the best-fit models), or $a'=0.002\pm 0.012$ and $b'=-0.26\pm 0.03$ (for the median models), with respect to $M_{\star{\rm e}}(K=0)$. The above results, with respect to $M_{\star{\rm e}}^{\rm A3D}$, are consistent with $a'=0$, again confirming the consistency between our results for $K=0$ and the \atl results. Regarding the slope $b'$ the \LCDM results give $b' \approx -0.20$, while the MOND results give $\approx -0.05$~dex shifted values with larger uncertainties.

 We use the median relations to correct the parameter $a$ of the FMP when $K\ne 0$ while keeping $b$ and $c$ fixed:
\begin{equation}
   a(K)=a(K=0)+b' K
 \label{eq:aK}  
\end{equation}
where $b'=-0.18 \pm 0.02$ ($\Lambda$CDM) or $-0.23 \pm 0.03$ (MOND). There are always some trade-offs among $a$, $b$, and $c$ when fitting the plane for real (imperfect) data. Hence, whether fixing $b$ and $c$ at the values for $K=0$ or allowing them to be free makes little difference about the plane for $K\ne 0$. For the sake of simplicity we fix $b$ and $c$ while allowing only $a$ to vary with $K$.

\begin{figure}[h] 
\begin{center}
\setlength{\unitlength}{1cm}
\begin{picture}(7,7)(0,0)
\put(-2.8,8.){\includegraphics{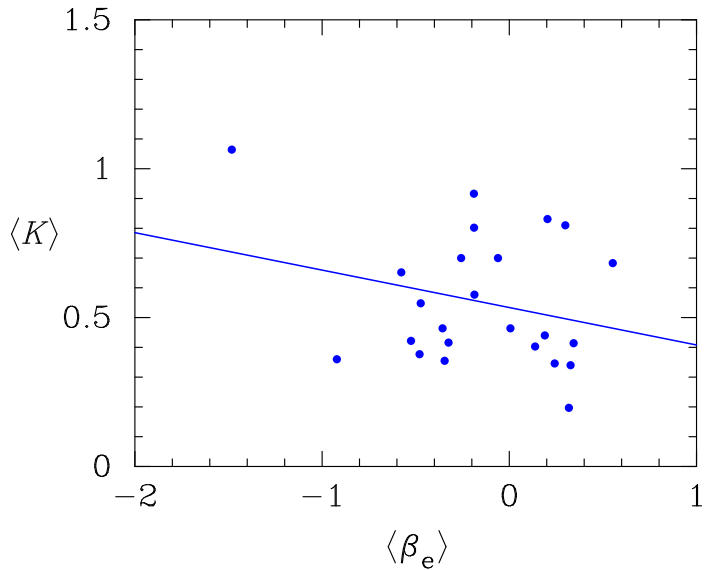}}
\end{picture}
\caption{Anti-correlation between median $M_\star/L$ gradient strength, $\langle K\rangle$, and median VD anisotropy, $\langle\beta_{\rm e}\rangle$, from the Monte Carlo sets of models of the \atl galaxies.}
\label{Kbete}
\end{center}
\end{figure}

The above results suggest that dynamical estimates of stellar masses with the \cite{vD17} $M_\star/L$ gradient would be $\approx -0.2$~dex shifted from that with constant $M_\star/L$. This shift is $\approx 0.1$~dex lesser than an estimate with isotropic velocity dispersions for all galaxies \citep{Ber18}. This is because in fitting the observed VDPs galaxies with larger gradients, $\langle K \rangle$, requires lower values of VD anisotropy $\langle \beta_{\rm e} \rangle$ (Figure~\ref{Kbete}), where $\beta_{\rm e}$ is the radially averaged anisotropy within $R_{\rm e}$, i.e.,
\begin{equation}
  \beta_{\rm e}\equiv \frac{1}{R_{\rm e}}\int_0^{\rm R_{\rm e}}\beta(r)dr.
  \label{eq:bete} 
  \end{equation}
This systematic variation of the fitted $\langle \beta_{\rm e} \rangle$ with $\langle K \rangle$ had a systematic impact on the scaling of Equation~(\ref{eq:aK}), weakening the slope $b'$ because the predicted VDP at fixed $K$ depends on VD anisotropy, as illustrated in Figure~2 of \cite{Ber18} using radially constant anisotropies.

\subsection{For the SDSS Galaxies}
 
For each SDSS galaxy with a measured geometric property (i.e.\ S\'{e}rsic parameters, $n$ and $R_{\rm e}$) and kinematic property (i.e.\ aperture VD, $\sigma_{\rm ap}$) we start by generating 90 Monte Carlo sets of $(R_{\rm e},n,w=\log_{10}\mu_{\rm \star e})$, where $\mu_{\rm \star e}=M_{\rm \star e}(K=0)/(\pi R_{\rm e}^2)$ is the projected stellar mass density when there is no $M_\star/L$ gradient. Parameters $R_{\rm e}$ and $n$ are generated from a bi-variate Gaussian distribution with their measurement errors and  a correlation coefficient of $\rho_{R_{\rm e},n}=0.07$ \citep{Mee13}. Parameter $w$ is generated from a Gaussian distribution with its mean and scatter estimated as follows. We first estimate $\sigma_{\rm e}$ using $\sigma_{\rm ap}$ through an empirical radial scaling:
\begin{equation}
\frac{\sigma_{\rm e}}{\sigma_{\rm ap}} = \frac{\langle\sigma_{\rm los}\rangle(R=R_{\rm e})}{\langle\sigma_{\rm los}\rangle(R=R_{\rm ap})} =  \left(\frac{R_{\rm e}}{R_{\rm ap}} \right)^\eta
  \label{eq:VPeta}
\end{equation}
where $\langle\sigma_{\rm los}\rangle(R)$ is given by Equation~(\ref{eq:VDap}) and we take $\eta=-[0.0392+0.0132(n-4)-0.0014(n-4)^2]$, as appropriate for bulge-dominated galaxies \citep{Ber17}, with an rms scatter of $0.03$. Then we estimate the mean value of $w$ using Equation~(\ref{eq:FMP}) with $a(K=0)$ and its uncertainty by adding all uncertainties of parameters and coefficients quadratically (i.e., assuming normal errors). (We also assign a central black hole using the recent result \citep{Sag} and a central core as described in \cite{CBK}, which are not however critical for our analysis, as $0.2R_{\rm e}\lesssim R_{\rm ap} \lesssim 1.2 R_{\rm e}$ for our selected SDSS galaxies.)

Given each set of $(R_{\rm e},n,w)$ for the galaxy under consideration, we generate a trial random set of $M_\star/L$ gradient strength $K$, VD anisotropy $\beta(r)$, and DM halo parameters $(\log_{10}M_{200},c_{\rm NFW},\alpha)$ under the \LCDM paradigm. Here, $\log_{10}M_{200}$ and $c_{\rm NFW}$ are assigned using the empirical relations (Section 2.5), but $\alpha$, $K$, and $\beta(r)$ are unknowns. For $\alpha$ we take a uniform deviate from the range $0< \alpha < 1.8$ as in modeling the \atl galaxies. For $K$ we take a uniform deviate from the range $0\le K  < K_{\rm max}$. Based on the modeling results for the \atl galaxies presented in \S 3.1 we choose $K_{\rm max}=1$ as our standard choice, and the median becomes $\approx 0.5$. We consider varying $K_{\rm max}$ from this standard value to estimate systematic uncertainties. Once a value of $K$ is assigned, the stellar mass based on Equation~(\ref{eq:FMP}) is shifted using Equation~(\ref{eq:aK}). For VD anisotropies we take the results of radially averaged values within $R_{\rm e}$ (Equation~\ref{eq:bete}) for the 24 pure-bulge \atl galaxies: $\langle\beta_{\rm e}\rangle=0.05\pm 0.25$, which is consistent with the literature \citep{Ger,Cap07,Tho}. Specifically, we take a median from the range $-0.2 \le \langle \beta_{\rm e}\rangle \le 0.3$ and take a Gaussian deviate of $\ln (\sigma_{\rm t}^2/\sigma_{\rm r}^2)_{\rm e} = \ln (1 -\beta_{\rm e})$ with a median $\ln(1-\langle\beta_{\rm e}\rangle)$ and standard deviation of $0.5$. We also consider radially varying anisotropies using the gOM model (Equation~\ref{eq:gOM}) with $r_a \le R_{\rm e}$ to study possible systematic uncertainties.

We then check if the aperture VD $\langle\sigma_{\rm los}\rangle(R=R_{\rm ap})$ (Equation~\ref{eq:VDap}) predicted for the trial set of parameters satisfies the measured SDSS VD $\sigma_{\rm ap}$. If $\sigma_{\rm ap}$ is matched within the given error, the set is taken. If not, another set is tried and repeated up to a predefined maximum number of iterations $N^{(1)}_{\rm max}=20$. If $\sigma(R_{\rm ap})$ is still not matched within the formal error, we then check whether $\sigma(R_{\rm ap})$ is matched within twice the formal error, iterating up to a larger maximum number $N^{(2)}_{\rm max}=50$. After $N^{(2)}_{\rm max}$ iterations, any random set is taken. Even if larger values of $N^{(1)}_{\rm max}$ and $N^{(2)}_{\rm max}$ than these choices are considered, the results are little changed.

 For the case of the MOND paradigm we use the general IF model given by Equation~(\ref{eq:IFnu}) and take for $a_0$ and $\nu$ uniform deviates, respectively, from $(0.5,\hspace{1ex}1.9)$ in units of $10^{-10}$~m~s$^{-2}$ and $(0.1,\hspace{1ex}2)$, which encompass all likely possibilities. 
  
 Assuming normal errors of all the measured and derived quantities, we would expect 68\% of the models in a Monte Carlo set to satisfy $\sigma_{\rm ap}$ within its uncertainty. We find that modeling with $K_{\rm max}=0$ (no $M_\star/L$ gradient) is successful in this sense for $\approx 89$\%  of galaxies. The success rate increases up to $\approx 99$\% if the $M_\star/L$ gradient ($K_{\rm max}> 0$) is considered. The failure rate of $\approx 1$\% -- $11$\% is reasonable, as the spherical approximation may not be good for some of our galaxies. A range of results can be produced by varying $K_{\rm max}$ and $\langle\beta_{\rm e}\rangle$ from the above specified ranges, but, the results also depend on the scaling of dynamical stellar mass, or parameter $b'$ in Equation~\ref{eq:aK}, on $K$. For our SDSS galaxies this scaling is necessary because dynamical stellar masses are assigned using the FMP rather than determined by the observed VDPs, as for the \atl galaxies. Thus, a Monte Carlo result is dictated primarily by the input choice of $(b',K_{\rm max},\langle\beta_{\rm e}\rangle)$. Based on the results for the \atl galaxies shown in Figures~\ref{MeK_A3D} and \ref{MeMDK_A3D}, our standard choice of $b'$ is $-0.18$ (the median result in Figure~\ref{MeK_A3D}). We consider $b'=-0.16$ and $-0.25$ to estimate systematic uncertainties.

 The Monte Carlo set of models produced for a given set of $(b',K_{\rm max},\langle\beta_{\rm e}\rangle)$ predicts a distribution of VP slope $\eta$ (Equation~\ref{eq:VPeta}) (see \cite{CG} for a previous example). We then check whether the prediction is consistent with the observed distribution of $\eta$ from the similarly selected 24 pure-bulge \atl galaxies: the mean  $\langle\eta\rangle = -0.055\pm 0.06$ and an rms scatter $s_\eta = 0.026$, which are derived by $\eta = \log(\sigma_e/\sigma_{e/8})/\log[R_{\rm e}/(R_{\rm e}/8)]$ taking the \atl reported $\sigma_e$ \citep{Cap13a} and $\sigma_{e/8}$ \citep{Cap13b} values. The required $\langle\beta_{\rm e}\rangle$ to be consistent with $\langle\eta\rangle= -0.055\pm 0.06$ depends on $K_{\rm max}$ and $b'$. The stronger the $K_{\rm max}$ or $b'$, the lower the $\langle\beta_{\rm e}\rangle$. For $K_{\rm max}=0$, $\langle\beta_{\rm e}\rangle>0$, but for $K_{\rm max}=1$ or larger, $\langle\beta_{\rm e}\rangle<0$. For a given $K_{\rm max}$, $b'=-0.25$ requires a lower $\langle\beta_{\rm e}\rangle$ compared with the standard choice $b'=-0.18$.  For any choice of the set $(b',K_{\rm max})$ from $b'=[-0.25,-0.16]$ and $K_{\rm max}=[0,1.5]$, respectively, there exists a value of $\langle\beta_{\rm e}\rangle$ in the range $[-0.2,0.3]$, so $\langle\eta\rangle = -0.055\pm 0.06$ is matched within $2\sigma$.

\begin{figure}[h] 
\begin{center}
\setlength{\unitlength}{1cm}
\begin{picture}(8,8)(0,0)
\put(-3.,9.3){\includegraphics{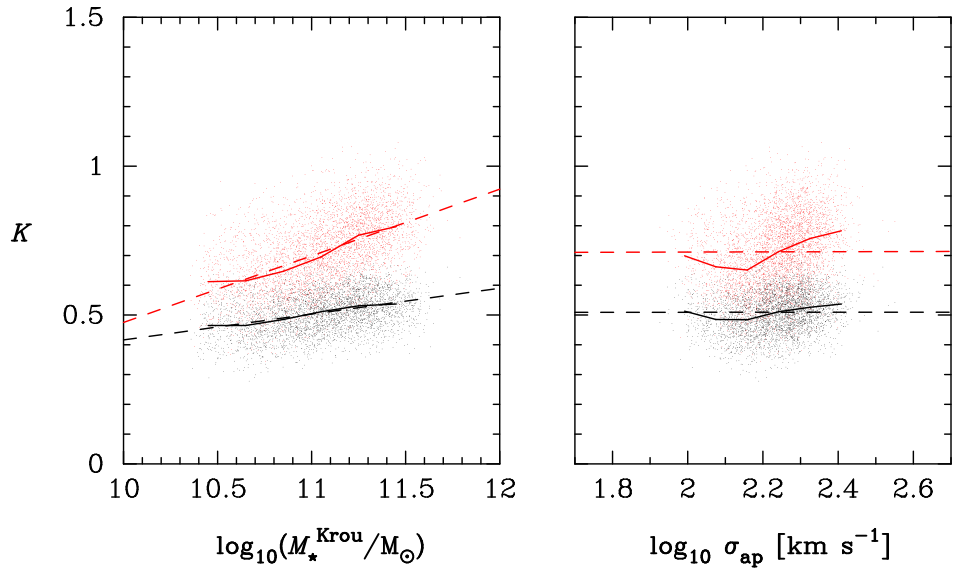}}
\end{picture}
\caption{Posterior distribution of $M_\star/L$ gradient strength $K$ with stellar mass $M_\star^{\rm Krou}$ (mass based on the Kroupa IMF and $K=0$) and SDSS aperture VD $\sigma_{\rm ap}$ from the Monte Carlo sets of models for $\sim 4000$ SDSS galaxies. The black and red points represent, respectively, the results with $K_{\rm max}=1$ and $1.5$. The solid curves represent medians, while dashed lines are least-squares fit results. These results indicate a correlation of $K$ with stellar mass but not with VD. }
\label{Kval}
\end{center}
\end{figure}

 The large number of SDSS galaxies allows us to investigate scalings of various quantities with, e.g., stellar mass or VD. Figure~\ref{Kval} exhibits the distribution of the $M_\star/L$ gradient strength $K$ with stellar mass and VD from the Monte Carlo sets with $K_{\rm max}=1$ and $1.5$.  The results indicate a correlation with stellar mass but not with VD. The correlation of $K$ with $\log_{10}(M_\star^{\rm Krou}/{\rm M}_\odot)$ depends on the assumed prior range of $K$. For $K_{\rm max}=1$, we find a relatively weak but significant slope as $K=(0.503\pm 0.002)+(0.087\pm 0.023)\left[\log_{10}\left(M_\star^{\rm Krou}/{\rm M}_\odot\right)-11\right]$. For $K_{\rm max}=1.5$, we have a stronger slope as  $K=(0.700\pm 0.002)+(0.224\pm 0.008)\left[\log_{10}\left(M_\star^{\rm Krou}/{\rm M}_\odot\right)-11\right]$. Note here that SDSS data alone cannot prefer one over the other because $K$ is not well constrained by the aperture VDs. However, the modeling results for the \atl galaxies give $\langle K\rangle =0.53^{+0.05}_{-0.04}$. Hence, a mild variation of $K$ with stellar mass such as the results with $K_{\rm max}=1$ is more likely. A positive correlation such as this would be in qualitative agreement with \cite{Par}. 

\section{Summary}

We have considered nearly round pure-bulge galaxies (24 \atl and 4201 SDSS galaxies) that are clearly distinct from rotating galaxies in structure and dynamics. We have carried out spherical Jeans modeling with radially varying VD anisotropies and radially varying $M_\star/L$ gradients with multiple goals: (1) to provide constraints on theories of DM and gravity through radial acceleration relation, independent of rotating galaxies; and (2) to investigate the galactic structure of pure-bulges with the $M_\star/L$ radial gradient. We have also presented simple analytical prescriptions to accommodate the $M_\star/L$ radial gradient using the MGE and S\'{e}rsic light distributions.

For the \atl galaxies the observed VD profiles are successfully reproduced with radially varying anisotropies (Equation~\ref{eq:gOM}) and a range of gradient strengths $K$ (Equation~\ref{eq:MLgrad}). We have obtained Monte Carlo sets of models under both the \LCDM and the MOND paradigms. Using these models we obtain the following results regarding dynamical stellar masses:
\begin{enumerate}
\item The FMP at fixed $K$ is consistent with the virial expectation, i.e.\ dynamical stellar mass $M_{\rm \star e} \propto \sigma_{\rm e}^2 R_{\rm e}$.
\item Dynamical stellar mass has the following scaling with $K$: $\log_{10} M_{\star{\rm e}}(K)\approx \log_{10} M_{\star{\rm e}}(K=0) - 0.2 K $.
\item The posterior distribution of $K$ varies from galaxy to galaxy without a tendency for a universal value. The distribution has a median of $\langle K\rangle = 0.53^{+0.05}_{-0.04}$ and an rms scatter of $s_{K}=0.17^{+0.06}_{-0.03}$. This is weaker than $K=1$ (the \cite{vD17} gradient), but in good agreement with the {\bf Salp$^{\rm In}$--Chab$^{\rm Out}$} gradient $K=0.555$ \citep{Ber18}. 
\item From the above results on $K$ and the scaling of the dynamical stellar mass with $K$ (Equation~\ref{eq:aK}), dynamical stellar masses, ignoring the $M_\star/L$ gradient are typically $\approx 0.1$~dex overestimated.
\end{enumerate}

We have also obtained Monte Carlo sets of models for the SDSS galaxies based on the accurate SDSS aperture VDs and the reliable FMP as a function of $K$ derived for the \atl galaxies. These Monte Carlo sets are useful for various statistical analyses, due to the large number of galaxies. From these sets we find:
\begin{enumerate}
\item There is an indication that $M_\star/L$ gradient has a positive correlation with stellar mass. The correlation starts to become significant if $\langle K\rangle \gtrsim 0.5$.
\end{enumerate}

We use the Monte Carlo sets of models for the \atl and SDSS galaxies obtained here to investigate the radial acceleration relation in \cite{CBS17}. Mass profiles, VD anisotropies, DM contents, or MOND IFs are analyzed in forthcoming papers.

\acknowledgments

{\bf Acknowledgments}: We thank Michele Cappellari for useful communications regarding the \atl data and modeling results. We also would like to thank the anonymous referee for the comments that helped us correct for errors and improve the presentation. This work was carried out at the University of Pennsylvania while K.-H.C. was on sabbatical leave. This research was supported by Basic Science Research Program through the National Research Foundation of Korea (NRF) funded by the Ministry of Education (NRF-2016R1D1A1B03935804).


\clearpage


\begin{thebibliography}{}

\bibitem[Abazajian et al.(2009)]{DR7} Abazajian, K. N., Adelman-McCarthy, J.~K., Ag\"{u}eros, M.~A., et al.\ 2009, \apjs, 182, 543

\bibitem[Bernardi et al.(2018a)]{Ber18} Bernardi, M., Sheth, R. K., Dominguez-Sanchez, H., et al.\ 2018a, MNRAS, 477, 2560
  
\bibitem[Bernardi et al.(2018b)]{Ber17} Bernardi, M., Sheth, R. K., Fischer, J.-L., et al.\ 2018b, MNRAS, 475, 757

\bibitem[Binney \& Mamon(1982)]{BM82} Binney, J, Mamon, G.~A. 1982, \mnras, 200, 361

\bibitem[Binney \& Tremaine(2008)]{BT} Binney, J., Tremaine, S. 2008, Galactic Dynamics, (2nd ed.; Princeton, NJ: Princeton Univ.\ Press)

\bibitem[Cappellari(2016)]{Cap16} Cappellari, M. 2016, \araa, 54, 597  

\bibitem[Cappellari et al.(2007)]{Cap07} Cappellari, M., Emsellem, E., Bacon, R., et al.\ 2007, \mnras, 379, 418

\bibitem[Cappellari et al.(2011)]{Cap11} Cappellari, M., Emsellem, E., Krajnovi\'{c}, D., et al.\ 2011, \mnras, 413,
 813

\bibitem[Cappellari et al.(2013a)]{Cap13a} Cappellari, M., Scott, N., Alatalo, K., et al.\ 2013a, \mnras, 432, 1709

\bibitem[Cappellari et al.(2013b)]{Cap13b} Cappellari, M., McDermid, R. M., Alatalo, K., et al.\ 2013b, \mnras, 432, 1862

\bibitem[Chabrier(2003)]{Chab} Chabrier, G. 2003, \pasp, 115, 763

\bibitem[Chae, Bernardi \& Kravtsov(2014)]{CBK} Chae, K.-H., Bernardi, M., Kravtsov, A. V. 2014, \mnras, 437, 3670 
  
\bibitem[Chae, Bernardi \& Sheth(2018)]{CBS17} Chae, K.-H., Bernardi, M., Sheth,  R. K. 2018, PhRvL, submitted  (arXiv:1707.08280v2)
  
\bibitem[Chae et al.(2012)]{Chae12} Chae, K.-H., Kravtsov, A. V., Frieman, J. A., Bernardi, M. 2012, JCAP, 11, 004

\bibitem[Chae \& Gong(2015)]{CG} Chae, K.-H., Gong, I.-T. 2015, \mnras, 451, 1719

\bibitem[Diemer \& Kravtsov(2015)]{DK} Diemer, B., Kravtsov, A. V. 2015, \apj, 799, 108

\bibitem[Djorgovski \& Davis(1987)]{DD} Djorgovski, S., Davis, M. 1987, \apj, 313, 59    

\bibitem[Dressler et al.(1987)]{Dre} Dressler, A., Lynden-Bell, D., Burstein, D.,  et al.\ 1987, \apj, 313, 42   

\bibitem[Emsellem, Monnet \& Bacon (1994)]{EMB} Emsellem, E., Monnet, G., \& Bacon, R. 1994, \aap, 285, 723 

\bibitem[Emsellem et al.(2011)]{Ems} Emsellem, E., Cappellari, M., Krajnovi\'{c}, D., et al.\ 2011, \mnras, 414, 888 
  
\bibitem[Famaey \& Binney(2005)]{FB} Famaey, B., Binney, J. 2005, \mnras, 363, 603

\bibitem[Famaey, Khoury \& Penco(2018)]{FKP} Famaey, B., Khoury, J., Penco, R. 2018, JCAP, 03, 038
  
\bibitem[Famaey \& McGaugh(2012)]{FM} Famaey, B., McGaugh, S. S. 2012, LRR, 15, 10

\bibitem[Gerhard et al.(2001)]{Ger} Gerhard, O., Kronawitter, A., Saglia, R. P., Bender, R. 2001, \aj, 121, 1936

\bibitem[Huertas-Company et al.(2001)]{HC} Huertas-Company, M., Aguerri, J. A. L., Bernardi, M., Mei,  S. , S\'{a}nchez Almeida, J. 2001, \aap, 525, 157 

\bibitem[Kent(1987)]{Ken} Kent, S. M. 1987, \aj, 93, 816

\bibitem[Krajnovi\'{c} et al.(2013)]{Kra} Krajnovi\'{c}, D., Alatalo, K., Blitz, L., et al.\ 2013, \mnras, 432,  1768

\bibitem[Kroupa(2002)]{Krou} Kroupa, P. 2002, Science, 295, 82
  
\bibitem[Lelli et al.(2017)]{Lel} Lelli, F., McGaugh, S. S., Schombert, J. M., Pawlowski, M. S. 2017, \apj, 836, 152

\bibitem[Mandelbaum, Seljak \& Hirata(2008)]{Man08} Mandelbaum, R., Seljak, U., Hirata, C. M. 2008, JCAP, 08, 006

\bibitem[Mandelbaum et al.(2016)]{Man16} Mandelbaum, R., Wang, W., Zu, Y., et al.\ 2016, \mnras, 457, 3200 

\bibitem[Mart\'{i}n-Navarro et al.(2015)]{MN} Mart\'{i}n-Navarro, I., La Barbera, F., Vazdekis, A., Falc\'{o}n-Barroso, J., Ferreras, I. 2015, \mnras, 447, 1033

\bibitem[Navarro, Frenk \& White(1997)]{NFW} Navarro, J. F., Frenk, C. S., White, S. D. M. 1997, \apj, 490, 493 

\bibitem[McGaugh(2004)]{McG04} McGaugh, S. S. 2004, \apj, 609, 652 

\bibitem[McGaugh(2005)]{McG05} McGaugh, S. S. 2005, \apj, 632, 859 

\bibitem[McGaugh, Lelli \& Schombert(2016)]{MLS} McGaugh, S. S., Lelli, F., Schombert, J. M. 2016, \prl, 117, 201101

\bibitem[Meert, Vikram \& Bernardi(2013)]{Mee13} Meert, A., Vikram, V., Bernardi, M. 2013, \mnras, 433, 1344

\bibitem[Meert, Vikram \& Bernardi(2015)]{Mee} Meert, A., Vikram, V., Bernardi, M. 2015, \mnras, 446, 3943

\bibitem[Merritt(1985)]{Merr} Merritt, D. 1985, \aj, 90, 1027

\bibitem[Milgrom(1983)]{Mil} Milgrom, M. 1983, \apj, 270, 371

\bibitem[Mo, van den Bosch \& White(2010)]{Mo10} Mo, H., van den Bosch, F. C., White, S. 2010, Galaxy Formation and Evolution (Cambridge: Cambridge Univ.\ Press)

\bibitem[More et al.(2011)]{More} More, S., van den Bosch, F. C., Cacciato, M., Skibba, R., Mo, H. J., Yang, X. 2011, \mnras, 410, 210 

\bibitem[Osipkov(1979)]{Osip} Osipkov, L. P. 1979, Pis'ma v Astron. Zhur., 5, 77
  
\bibitem[Parikh et al.(2018)]{Par} Parikh, T., Thomas, D., Maraston, C., et al.\ 2018, \mnras, 477, 3954

\bibitem[Saglia et al.(2016)]{Sag} Saglia, R. P., Opitsch, M., Erwin, P., et al.\ 2016, \apj, 818, 47

\bibitem[Salpeter(1955)]{Salp} Salpeter, E. E. 1955, \apj, 121, 161

\bibitem[Scott et al.(2013)]{Scot} Scott, N., Cappellari, M., Davies, R. L., et al.\ 2013, \mnras, 432, 1894

\bibitem[S\'{e}rsic(1968)]{Ser} S\'{e}rsic, J. L. 1968, Atlas de Galaxias Australes (C\'{o}rdoba: Observatorio Astron\'{o}mico)

\bibitem[Sonnenfeld et al.(2018)]{Son} Sonnenfeld, A., Leauthaud, A.,  Auger, M. W., et al.\ 2018, \mnras, submitted (arXiv:1801.01883) 
  
\bibitem[Thomas et al.(2013)]{ThoD} Thomas, D., Steele, O., Maraston, C., et al.\ 2013, \mnras, 431, 1383 

\bibitem[Thomas et al.(2007)]{Tho} Thomas, J., Saglia, R. P., Bender, R., et al.\ 2007, \mnras, 382, 657

\bibitem[van Dokkum et al.(2017)]{vD17} van Dokkum, P., Conroy, C., Villaume, A. , Brodie, J., Romanowsky, A. J. 2017, \apj, 841, 68 

\end{thebibliography}
\end{document}